\documentclass[fleqn,usenatbib]{mnras}

\usepackage{newtxtext,newtxmath}

\usepackage[T1]{fontenc}
\usepackage{ae,aecompl}

\usepackage{graphicx}	
\usepackage{caption}
\usepackage{amsmath}    
\usepackage{bm}
\usepackage{amsfonts}
\usepackage{subcaption}
\usepackage{float}
\usepackage{gensymb}

\title[UV bright stellar populations in NGC 1261]{Study of UV bright stellar populations in the globular cluster NGC 1261 using AstroSat}

\author[Rani et al.]{
Sharmila Rani,$^{1,2}$\thanks{E-mail: sharmila.rani@iiap.res.in}
Gajendra Pandey,$^{1}$
Annapurni Subramaniam,$^{1}$
Snehalata Sahu,$^{1}$
\newauthor N. Kameshwara Rao$^{1}$\\
$^{1}$Indian Institute of Astrophysics, Bangalore, 560034,  India\\
$^{2}$Pondicherry University, R.V. Nagar, Kalapet, 605014, Puducherry, India
}

\date{Accepted XXX. Received YYY; in original form ZZZ}
\pubyear{2020}

\begin{document}
\label{firstpage}
\pagerange{\pageref{firstpage}--\pageref{lastpage}}
\maketitle
\begin{abstract}
We present the UV photometry of the globular cluster NGC 1261 using images acquired with the Ultraviolet Imaging Telescope (UVIT) on-board ASTROSAT. We performed PSF photometry on four near-UV (NUV) and two far-UV (FUV) images and constructed UV colour-magnitude diagrams (CMDs), in combination with HST, Gaia, and ground-based optical photometry for member stars. We detected the full horizontal branch (HB) in NUV, blue HB in the FUV and identified two extreme HB (EHB) stars. HB stars have a tight sequence in UV-optical CMDs well fitted with isochrones generated (12.6 Gyr age, [Fe/H] = $-$1.27 metallicity) using updated BaSTI-IAC models. 
Effective temperatures ($T_{eff}$), luminosities and radii of bright HB stars were estimated using spectral energy distribution. As we detect the complete sample of UV bright HB stars, the hot end of the HB distribution is found to terminate at the G-jump ($T_{eff}$ $\sim$ 11500 K). The two EHB stars, fitted well with single spectra, have $T_{eff}$ = 31,000 K and a mass = 0.495M$_\odot$, and follow the same $T_{eff}$ - Radius relation of the blue HB stars.  We constrain the formation pathways of these EHB stars to extreme mass loss in the RGB phase (either due to rotation or enhanced Helium), OR early hot-flash scenario. 
\end{abstract}

\begin{keywords}
(Galaxy:) globular clusters: individual: NGC 1261 - stars: horizontal branch, (stars:) blue stragglers - (stars:) Hertzsprung-Russell and colour-magnitude diagrams
\end{keywords}



\section{Introduction}
Globular clusters (GCs) are ideal laboratories to study stellar evolution at different phases starting from main sequence (MS), sub-giant branch (SGB), red giant branch (RGB), horizontal branch (HB), asymptotic and post-asymptotic giant branch (PAGB), and finally the white dwarf (WD) phase. GCs are the oldest and the most massive stellar aggregates known to exist in our Galaxy, consisting of simple and multiple stellar populations \citep{2011MNRAS.415..643W, 2013ApJ...778..186V, 2004MSAIS...5..105B, 2007ApJ...661L..53P, 2008A&A...490..625M, 2008ApJ...673..241M, 2008ApJ...672L..29Y, 2009ApJ...697L..58A, 2009Natur.462..480L, 2009IAUS..258..233P, 2010sf2a.conf..319M, 2012ApJ...760...39P, 2015AJ....149...91P} with single stars, binary stars as well as multiple systems. Hence, they provide the best platform to study the exotic populations such as blue straggler stars (BSSs), cataclysmic variables, low mass X-ray binaries (LMXB) etc., which demand binary formation pathways. However, recent studies have shown that almost all the GCs consist of multiple stellar populations.
Ultraviolet (UV) study of evolved populations in GCs is very important because a few critical evolutionary phases, found in the GCs, are brighter in UV than in optical wavelength. These are the main contributors to the luminosity of GCs in the UV.  These UV bright stars are hot and mostly consists of HB stars, PAGB stars, WDs, and BSSs \citep{1972A&A....18..390Z, 1983PASP...95..256H}. The spectral energy distribution (SED) of these hot stars peaks at shorter wavelengths mainly in the near-ultraviolet (NUV) or far-ultraviolet (FUV). Note that MS stars and RGB stars, which make up the majority of stars in a GC, are faint at such short wavelengths. Hence, the UV images are less crowded than the optical images. 
Recent studies of GCs  \citep{2003ApJ...588..464F, 2010ApJ...710..332D, 2010ApJ...722..158H, 2016ApJ...830..139P,2017MNRAS.469..267D, 2017ApJ...839...64R} have highlighted the importance of UV CMDs to identify and study the properties of UV bright stellar populations. 
The HB stars are core helium burning low mass stars with hydrogen rich envelope surrounding the core. The core mass of stars lying on the HB is approximately $\sim0.5M_{\odot}$ (\citealp{1970ApJ...161..587I}). The HB itself consists of sub-populations separated by gaps, namely, the red horizontal branch (RHB), the blue horizontal branch (BHB) and the extreme horizontal branch (EHB). The RHB stars are cooler than the RR Lyrae instability strip with their effective temperatures ranging from $\sim5,000-6,200$ K whereas the BHB stars are hotter than the RR Lyrae instability strip and cooler than the EHB stars. The effective temperature of the BHB stars are expected to lie between $\sim8,000-20,000$ K. EHB stars are core helium burning stars with an envelope too thin to sustain hydrogen burning. These stars have effective temperature greater than 20,000 K and are expected to lie at the end of the blue tail of the HB in the optical CMD. UV photometry helps to confirm the presence of the EHB stars and separate them from bright BSSs, as optical colours become degenerate at the effective temperature of these stars \citep{2000ApJ...530..352D, 2001ApJ...562..368B, 2008ApJ...677.1069D}. Parameters and reasons behind the  distribution of stars along the HB in optical CMDs are not yet understood.
At first, it was thought that only metallicity of the cluster affected the HB morphology, but later through observations it was found that parameters other than metallicity, such as age of the cluster and/or He abundance, can affect the colour distribution of the HB stars (\citealp{2009Ap&SS.320..261C}). This problem is known as second parameter problem, first mentioned by Sandage \& Wallerstein \citep{1960ApJ...131..598S, 1967ApJ...150..469S, 1967AJ.....72...70V}. \cite{2014ApJ...785...21M} investigated the relation between HB morphology and various properties in GCs. They suggested that age and metallicity are the main global parameters of the HB morphology of GCs, while the range of He abundance within a GC is the main nonglobal parameter.\\

In clusters with the HB morphology covering a wide range in optical colour, 
a number of discontinuities or jumps have been detected, although the visibility of these gaps change with the band passes utilized (\citealp{2016ApJ...822...44B}). 
Three prominent such discontinuities or gaps known so far are "Grundahl jump" (G-jump) which lies within the BHB at $\sim$ 11500 K (\citealp{1999ApJ...524..242G}), "Momany jump" (M-jump) lies within the EHB at $\sim$ 23000 K \citep{2002ApJ...576L..65M, 2004A&A...420..605M} and the gap between the EHB and the blue-hook stars, covering $\sim$ 32000-36000 K (\citealp{2001ApJ...562..368B}). The gaps in the HB distribution are discernible in clusters hosting sufficient number of BHB and EHB stars and these gaps are found to be consistent in colour (effective temperature) across the clusters \citep{1998ApJ...500..311F, 2016ApJ...822...44B}. Various explanations are given for the presence of these gaps in HB distribution by (\citealp{2001PASP..113.1162M}), such as diverging evolutionary paths, mass loss on RGB phase, distinctions in CNO or rotation rates, dynamical interactions, atmospheric processes, He mixing in RGB stars, and statistical fluctuations. 
\cite{2016ApJ...822...44B} characterized these HB features in 53 GCs, including NGC 1261. They created colour-colour plots for all selected clusters. They found less than four HB stars blue-ward of the G-jump in NGC 1261 (see Figure 5 in \citealp{2016ApJ...822...44B}). \cite{2012AJ....143..121S} constructed UV CMDs (FUV$-$NUV vs FUV) for 44 galactic globular clusters using GALEX data. They observed that the HB stars follow a diagonal sequence, unlike the horizontal distribution in optical and  its slope mainly depends on the bolometric correction effects.\\ 

EHB stars are intriguing objects to study as they are one of the important sources of the UV upturn at wavelengths shorter than 2300 {\AA} in the spectra of elliptical galaxies \citep{1990ApJ...364...35G,1995ApJ...442..105D,2011ApJ...740L..45C,2012ApJ...747...78B}. The origin of EHB stars in GCs is still not clear. One of the formation scenario suggested for EHB stars is enhanced mass loss on the RGB evolutionary phase. \cite{1976ApJ...204..488M} and \cite{1987fbs..conf..435T} proposed that the EHB stars are the product of dynamical interactions inside binary systems in clusters. He-mixing is also a possible scenario to explain the formation of EHB stars in clusters (\citealp{1997ApJ...474L..23S}). Through convection, helium is mixed into the outer layers of the envelope, which in turn, increase the envelope He abundance. The enhanced He abundance causes the tip of the RGB to have higher luminosity than in normal case. The higher luminosity leads to high mass loss on the RGB. Thus, He mixing is responsible for producing a bluer HB morphology. Other possible formation scenarios for EHB stars in clusters are early hot-flasher, late hot-flasher and He enrichment \citep{2001ApJ...562..368B,2010ApJ...718.1332B,2012ASPC..452...23B, 2015MNRAS.449.2741L, 2016PASP..128h2001H}.\\

  BSSs occupy a region above the turn-off (TO) point in the colour-magnitude diagram (CMD) of star clusters, where no stars are expected on the basis of standard stellar evolution, if we assume that all cluster stars are coeval. It has been suggested that BSSs are the product of stellar collisions or mass exchange in close binary systems \citep{1964MNRAS.128..147M, 1976ApL....17...87H}. Stellar collisions formation scenario for BSSs dominates in high-density environment whereas other formation scenarios dominate in a low-density environment.\\ 

Here, we study the southern galactic GC NGC 1261 located in constellation Horologium at a distance of 17.2 kpc (\citealp{2019RMxAA..55..337A}) and metallicity \big[Fe/H\big] = -1.27 dex (\citealp{2009A&A...508..695C}). Since this cluster is away from the galactic disc (b = $-52\degree.13$), it experiences almost negligible or no reddening. The age of this cluster is estimated to be 10.75 $\pm$ 0.25 Gyr (\citealp{2013ApJ...775..134V}), 11.5 $\pm$ 0.5 Gyr (\citealp{2010ApJ...708..698D}) and 12.6 $\pm$ 1 Gyr (\citealp{2013A&A...558A..53K}). In this study, the adopted age of this cluster is 12.6 Gyr to generate the isochrones. In this paper, we present the results of a UV imaging study of  NGC 1261 in six filters (2 FUV and 4 NUV), using the Ultra-violet Imaging Telescope (UVIT) on-board ASTROSAT. We used the proper motion estimated using HST and Gaia DR2 data to select cluster members in the inner and outer region of the cluster, respectively. We detect the HB and a few UV bright stars among the HB population. We determine the properties of the HB stars by analysing the SEDs to shed light on the formation and evolution of bright HB stars.\\


This paper is laid out as follows. Section~\ref{sec:2} describes the observations and data reduction. In Section~\ref{sec:3}, the UV and Optical CMDs are presented. We describe the properties of bright HB stars derived from the UVIT photometry along with HST, Gaia and Ground based photometry in Section~\ref{sec:4}. 
All the results are discussed in Section~\ref{sec:5}. Our results are summarised and concluded in Section~\ref{sec:6}.

\section{Observations and Data Reduction}
\label{sec:2}
NGC 1261 was observed with UVIT on 26 August 2017 in two FUV and four NUV filters. 
UVIT is one of the payloads in ASTROSAT, the first Indian Space Observatory. UVIT can perform simultaneous imaging in three channels: FUV(130-180nm), NUV(200-300nm) and VIS(320-550nm), in a circular field of view of diameter $\sim28{\arcmin}$. The spatial resolution (FWHM) is better than $1.8{\arcsec}$ for the FUV and NUV channels, and it is $2.2{\arcsec}$ for the VIS channel. The VIS channel is used only for drift correction. Each channel (FUV, NUV and VIS) consists of a set of selectable filters having different wavelength range. The intensified CMOS detectors are used in photon counting mode for the two UV channels and integration mode for the VIS channel. Further details about UVIT and calibration results can be found in \citep{2016SPIE.9905E..1FS, 2017JApA...38...28T}. The primary photometric calibration for all FUV and NUV channels were accomplished using observations of HZ4, a white dwarf spectrophotometric standard star (\citep{2017AJ....154..128T}). 
The magnitude system adopted for UVIT filters is the AB magnitude system, and hence the estimated magnitudes will be in this system.\\

\begin{figure}
    \centering
	\includegraphics[scale=0.4]{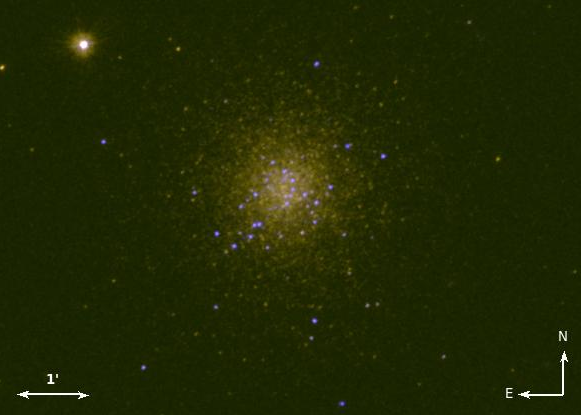}
    \caption{UVIT image of NGC 1261 obtained by combining images in NUV (N279N) and FUV (F172M) channels.}
    \label{fig:1}
\end{figure}

ASTROSAT has an orbit of 90 min., and UVIT observes only the night part of the orbit. The observation is, therefore, about 20 min or less per orbit (after the overheads). The UVIT instrument takes data over several orbits to complete the required exposure time. The images are created for each orbit using a customised software package, CCDLAB (\citealp{2017PASP..129k5002P}), by correcting for the geometric distortion, flat field illumination, and spacecraft drift. They are then aligned and combined to create a science ready image. The photometry is carried out on these final science ready images.
 The details of observations and photometry of NGC 1261 UVIT images are given in Table~\ref{tab:1}. The UVIT image created using one NUV N279N and one FUV F172M filter is shown in Figure~\ref{fig:1}. In this figure, blue and yellow colours represent UVIT FUV and NUV detections respectively. In the FUV, we are able to resolve stars up to the center of the cluster, but in the case of NUV images crowding is present in the central regions. Isolated stellar sources in the UVIT images have FWHM $\sim1.2{\arcsec}$ and $\sim1.5{\arcsec}$ in NUV and FUV channels, respectively.\\

\begin{figure}
   \hspace*{-0.7cm} 
	\includegraphics[height=9.5cm, width=10cm]{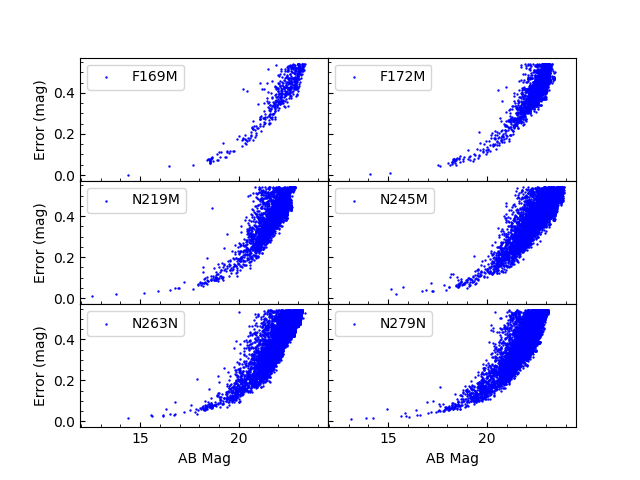}
     \caption{PSF fit errors as a function of magnitude for the UVIT observations of NGC 1261. The top two panels show the plot for two FUV filters, whereas the bottom 4 panels show for the NUV filters.}
    \label{fig:2}
\end{figure}

 The PSF photometry was performed on all the UVIT images using DAOPHOT package in IRAF (\citealp{1987PASP...99..191S}). The various steps to carry out crowded field photometry are following. First stars are identified in the images using DAOFIND package in DAOPHOT. Then, we computed the magnitudes of stars 
using DAOPHOT task PHOT. 
The model point spread function (PSF) was generated by choosing isolated and moderately bright stars.The generated model PSF is then applied to all the detected stars using ALLSTAR task in DAOPHOT to obtain the PSF fitted magnitudes. 
A curve-of-growth analysis technique was carried out to estimate aperture correction value in each filter and applied it to the PSF generated magnitudes. Finally, the saturation correction (For more details, see \citep{2017AJ....154..128T}) was done to obtain final magnitudes in each filter. The instrumental magnitudes have been calibrated to AB magnitude system by using zero-point magnitudes reported in the calibration paper \citep{2017AJ....154..128T}.  
The Figure~\ref{fig:2} shows the PSF fit error plots for all filters as a function of magnitude. We detect stars as faint as 21 magnitude in NUV and 22 magnitude in FUV with typical errors 0.2 mag and 0.3 mag, respectively.
 
\begin{table}
	\hspace{1.5 cm}
	\caption{Details of observations and photometry of NGC 1261}
	\label{tab:1}
	\begin{tabular}{cccccc} 
		\hline
		\hline
		 Filter & $\lambda_{mean}$ & $\Delta\lambda$  & ZP  & Exposure & No. of \\
		  & ({\AA}) & ({\AA}) & (AB mag) & Time (sec) & stars \\
		\hline
		 F169M & 1608 & 290 & 17.45 & 1746 & 486 \\
		 F172M & 1717 & 125 & 16.34 & 6662 & 2181 \\
		 N219M & 2196 & 270 & 16.59 & 2847 & 2610 \\
		 N245M & 2447 & 280 & 18.50 & 740 & 4755 \\
	 	 N263M & 2632 & 275 & 18.18 & 1022 & 6037\\
		 N279N & 2792 & 90  & 16.50 & 3831 & 9363 \\
		\hline
	\end{tabular}
\end{table}

\section{The UV and Optical Colour-Magnitude Diagrams}
\label{sec:3} 
\begin{figure}
   \hspace{-0.9cm} 
   \includegraphics[scale=0.5]{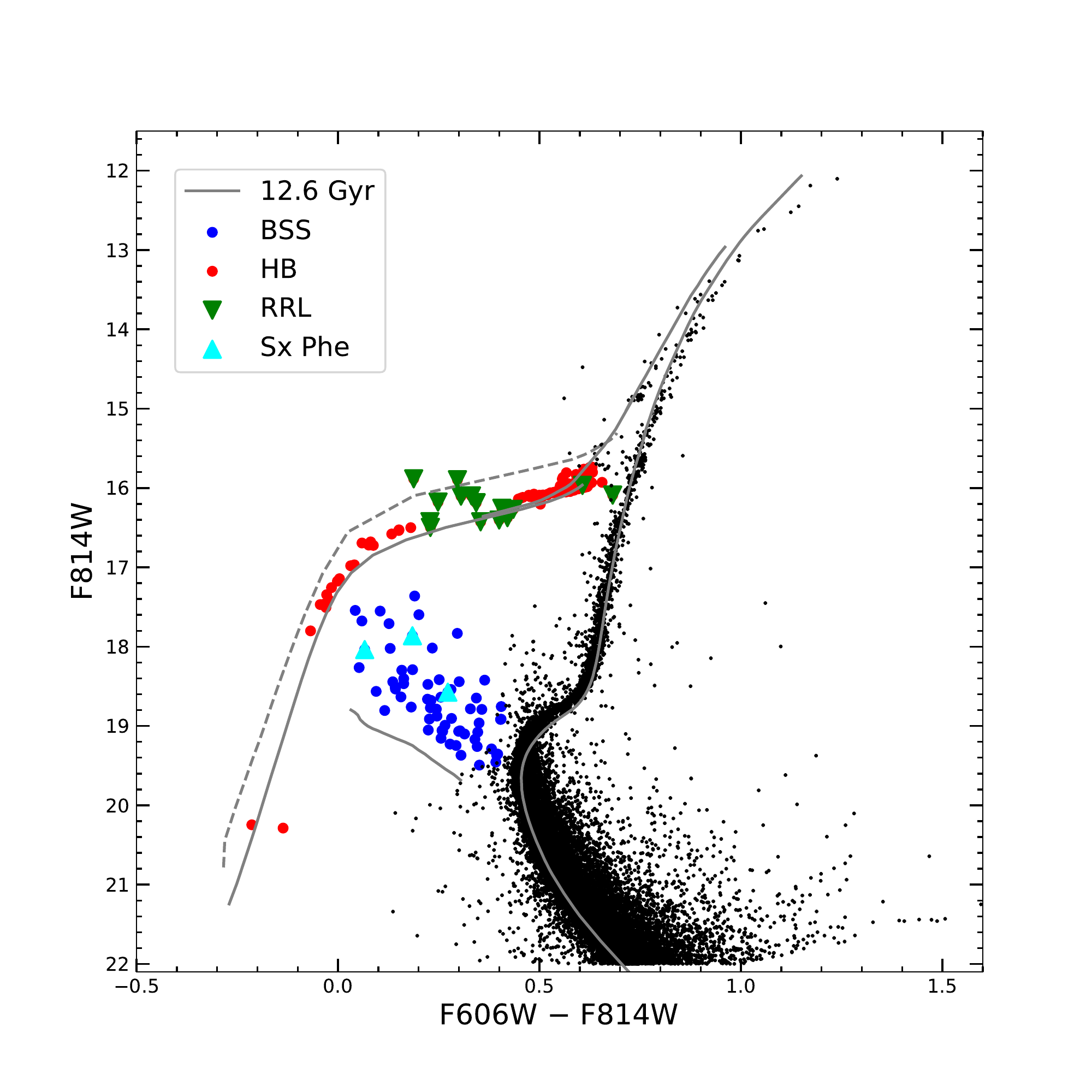}
     \caption{Optical CMD of NGC 1261 for the region covered by the HST.  Black dots represent the HST detected stars with proper motion membership probability more than $90\%$. only HB and BSS stars detected in the NUV N279N filter and cross-matched with HST catalog are shown in different colours, which are explained in the figure. The variable stars such as RR Lyrae and SX Phe are also shown in figure. The solid grey lines denote updated BaSTI-IAC model isochrone for 12.6 Gyr and \big[Fe/H\big] = -1.27 dex. The solid grey line on the HB locus is the zero-age HB (ZAHB) and the dotted one represents the terminal-age HB (TAHB), where the star has completed 99$\%$ of its He-burning lifetime. We also show the BSS model line which is extension of the zero age main sequence (ZAMS).}
    \label{opthstcmd}
\end{figure}

\begin{figure*}
   \hspace*{-1.4cm} 
\includegraphics[height=20cm,width=20cm]{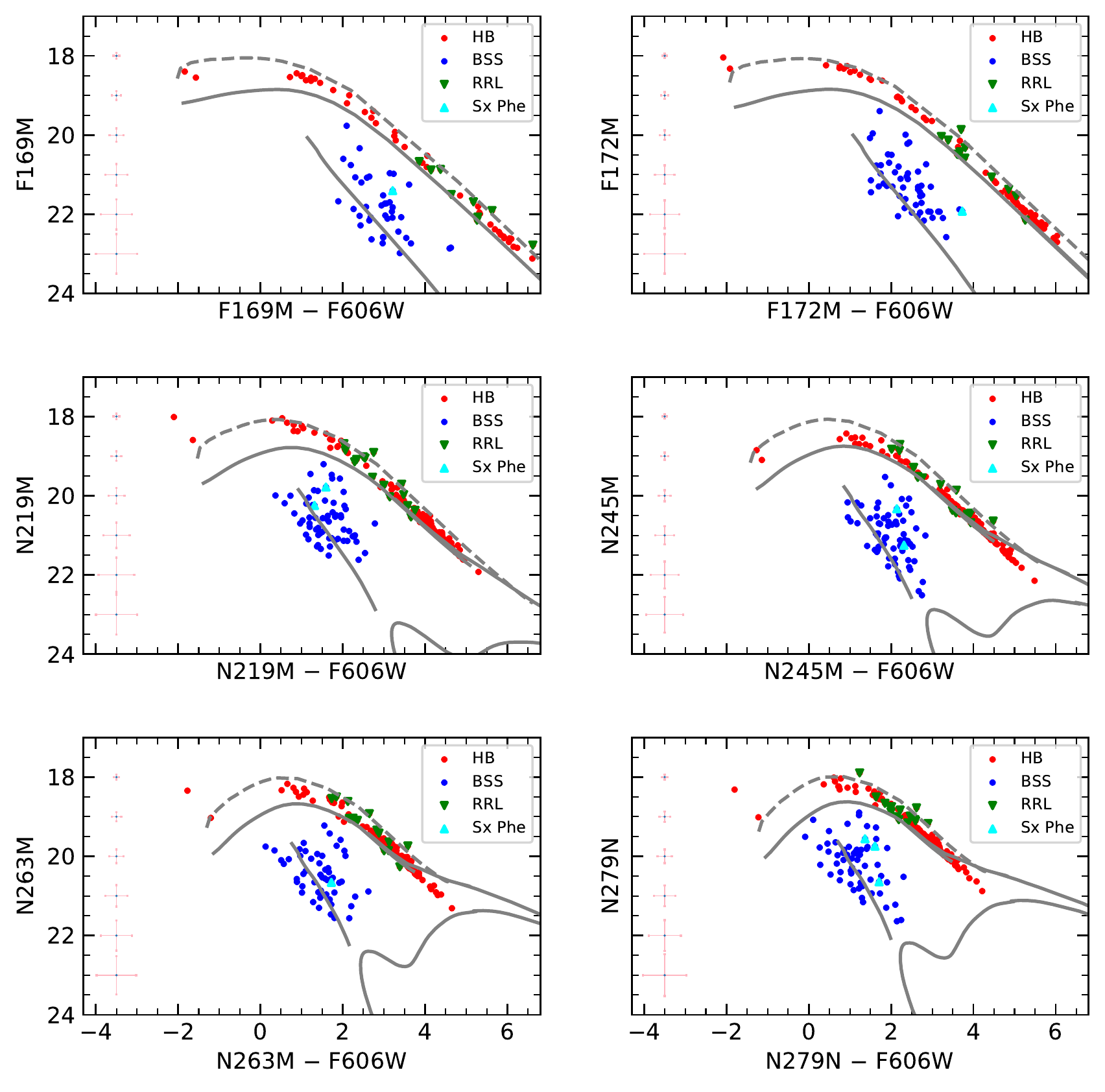}
     \caption{UV-optical CMDs of NGC 1261 after cross-matching HST data with UVIT data in 4 NUV and 2 FUV filters. The meaning of different colours and symbols is shown in the panels. The photometric errors in magnitude and colour are also shown in each panel. The grey lines denote the updated BaSTI-IAC model isochrone for 12.6 Gyr and \big[Fe/H\big] = $-$1.27 dex. The solid and dotted grey lines are ZAHB and TAHB, respectively.}
    \label{optuvhstcmds}
\end{figure*}

Colour-Magnitude Diagrams (CMDs) are a very important tool to identify different evolutionary sequences in star clusters. As we have observed this cluster in 4 NUV and 2 FUV filters of UVIT, a large number of CMDs using various filter combinations are possible. 
In order to identify stars in the inner region (within a diameter $\sim3.4'$), we have used HST UV legacy survey catalog of GCs (\citealp{2018MNRAS.481.3382N}) to cross-match with UVIT detected sources. Most likely members of this cluster in the inner region have proper motion probability more than $90\%$ as mentioned by \cite{2018MNRAS.481.3382N}. In order to cross-match UVIT detected sources with HST data, first we selected stars with proper motion membership probability more than $90\%$. In HST, the filters F606W and F814W are proxy to V and I bands. 
To cross match HB stars and BSSs, we have used specific colour and magnitude range in the optical HST CMD. In the F606W$-$F814W vs F814W HST CMD, we have used $-$0.1<F606W$-$F814W<0.7 and 15<F814W<20.5 range for HB stars and  0<F606W$-$F814W<0.4 and  17.3<F814W<19.5 range for BSSs. Two stars appearing at the faint end of the blue tail of HB in optical CMD shown in Figure~\ref{opthstcmd} are also present in the colour-colour plot created for this cluster by \cite{2016ApJ...822...44B} (see their Figure 5). The Vega magnitude system used in HST is converted into the AB magnitude system, in order to adopt the same magnitude system\footnote{\url{http://waps.cfa.harvard.edu/MIST/model_grids.html##zeropoints}}. The magnitude system used to create optical CMDs in the inner and outer region is the Vega magnitude system, but both photometric systems are different. \cite{2001AJ....122.2587C}, \cite{2016AJ....152...55S} and \cite{2019RMxAA..55..337A} have studied the variability of stars in  this cluster. To identify the variable stars such as RR Lyrae and Sx Phe, we have cross-matched UVIT data with variable star catalog from \cite{2019RMxAA..55..337A}. In NUV bands, our sample of HB stars is not complete in the inner region as we are unable to resolve stars in the inner $1'$ diameter region due to crowding whereas, outside this region, we detect about 90\% stars compared to the HST. The total number of detected HB, BSSs and variable stars in each UVIT filter is tabulated in Table~\ref{tab:2}. The optical and UV-optical CMDs are shown in Figure~\ref{opthstcmd} and \ref{optuvhstcmds}, respectively. The photometric error bars shown in Figures~\ref{optuvhstcmds} and \ref{optuvgaiacmds} are the median of the photometric errors of stars at a selected magnitude ranges. In the UV-optical CMDs, we observe that the HB stars no longer follow horizontal sequence as found in the optical CMDs, rather it follows a diagonal sequence with less spread \citep{2007A&A...474..105B, 2008ApJ...677.1069D, 2011MNRAS.410..694D, 2009MNRAS.394L..56D, 2010ApJ...710..332D,   2019MNRAS.482.1080S, 2017AJ....154..233S, 2017ApJ...839...64R}. However, note that the sequence of BSSs, if diagonal, remains unaltered in both UV-optical and optical CMDs.\\


 The optical and UV-optical CMDs are overlaid with updated BaSTI-IAC isochrones (\citealp{2018ApJ...856..125H}). The updated BaSTI-IAC\footnote{\url{http://basti-iac.oa-abruzzo.inaf.it/}} isochrones are generated for an age 12.6 Gyr (\citealp{2013A&A...558A..53K}), a distance modulus of 16.21 mag (\citealp{2019RMxAA..55..337A}) and a metallicity \big[Fe/H\big] = $-$1.27 dex (\citealp{2009A&A...508..695C}) with helium abundance Y = 0.247, \big[$\alpha$/H\big] = 0, including overshooting, diffusion, and mass loss efficiency parameter $\eta$ = 0.3. The BaSTI-IAC model also provides HB model, which includes zero age HB (ZAHB), post-ZAHB tracks and end of the He phase known as terminal age HB (TAHB) with or without diffusion for a particular mass range. We generated the ZAHB and TAHB tracks for a metallicity \big[Fe/H\big] = $-$1.27 dex including diffusion. The BaSTI-IAC model does not provide the BSS model line. In order to define the location of BSSs in CMDs, we used BaSTI isochrones \citep{2004ApJ...612..168P} generated using FSPS code of \cite{2009ApJ...699..486C}, \cite{2010ApJ...712..833C}. The observed UVIT stellar magnitudes are corrected for reddening and extinction. We have adopted a reddening E(B$-$V) of 0.01 mag from (\citealp{2013A&A...558A..53K}) and the ratio of total-to-selective extinction as $R_v$ = 3.1 from (\citealp{1958AJ.....63..201W}) for the Milky Way. The extinction co-efficient in visible is $A_V$ = 0.031 mag. The $A_V$ is used to determine extinction co-efficients $A_\lambda$ for all pass bands using the reddening relation of \cite{1989ApJ...345..245C}. The BS sequence shown in optical as well as in UV CMDs is the extension of the zero-age main sequence (ZAMS). It should roughly follow the ZAMS up to twice the turn-off mass if BSSs are the product of mergers between two MS stars. The MS turn-off mass in NGC 1261 is approximately 0.8M$_{\odot}$. The brightest and faintest part of the BS sequence shown in all CMDs corresponds to 1.6M$_{\odot}$ and 1.1M$_{\odot}$, respectively.\\
 \begin{figure}
 \hspace*{-0.9cm} 
\includegraphics[scale=0.5]{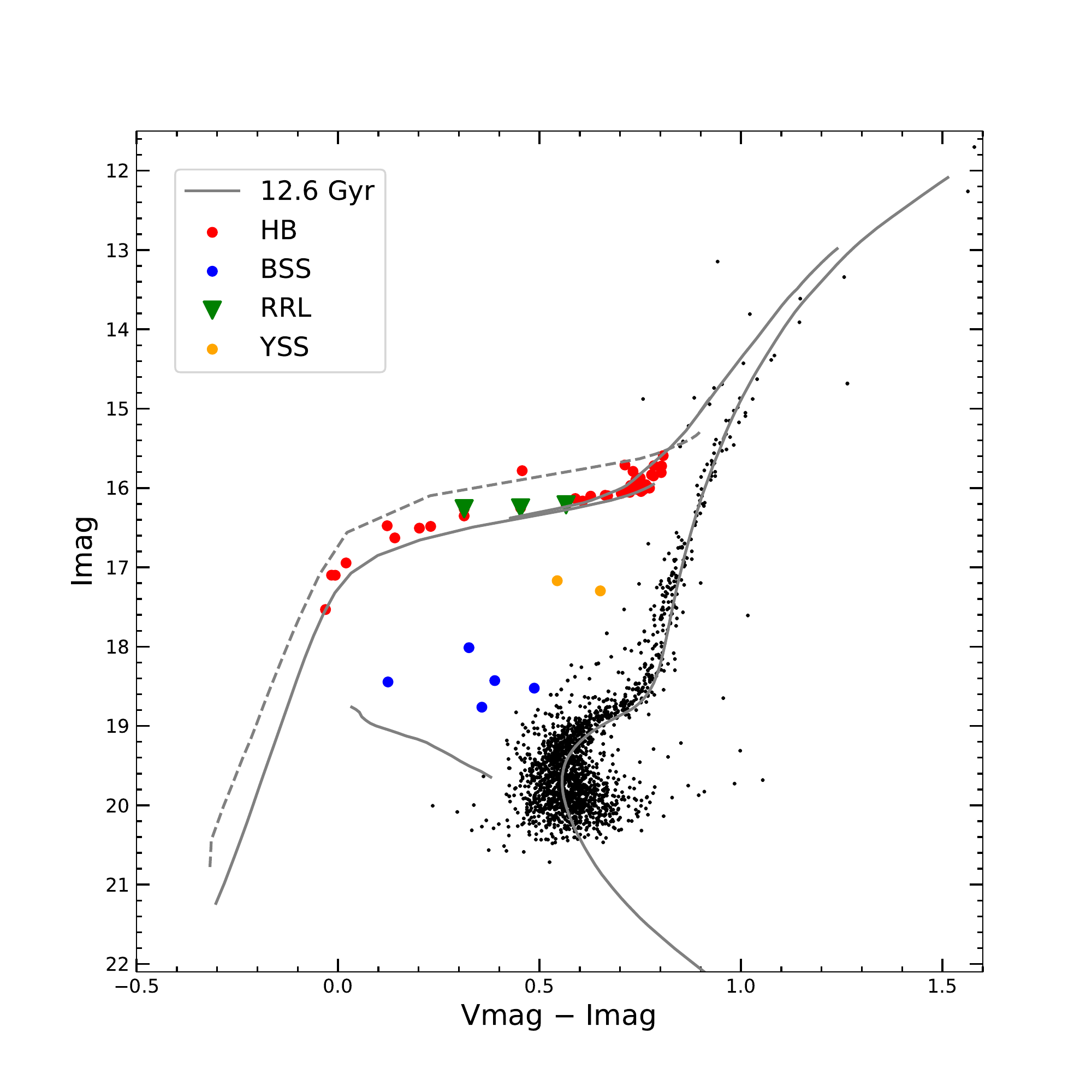}
     \caption{Optical CMD of NGC 1261 for the region outside the HST field. HB and BSS stars that are detected in the NUV N279M UVIT filter and cross-matched with ground based photometric data (\protect\citealp{2010A&A...516A..23K}) and Gaia data are marked with different colors. Rest of the stars shown with black dots are cross-matched ground based data with Gaia data. Other details are same as in Figure~\protect\ref{opthstcmd}.}
    \label{optgaiacmd}
\end{figure}
 
 \begin{figure*}
\hspace*{-1.4cm} 
\includegraphics[height=20cm,width=20cm]{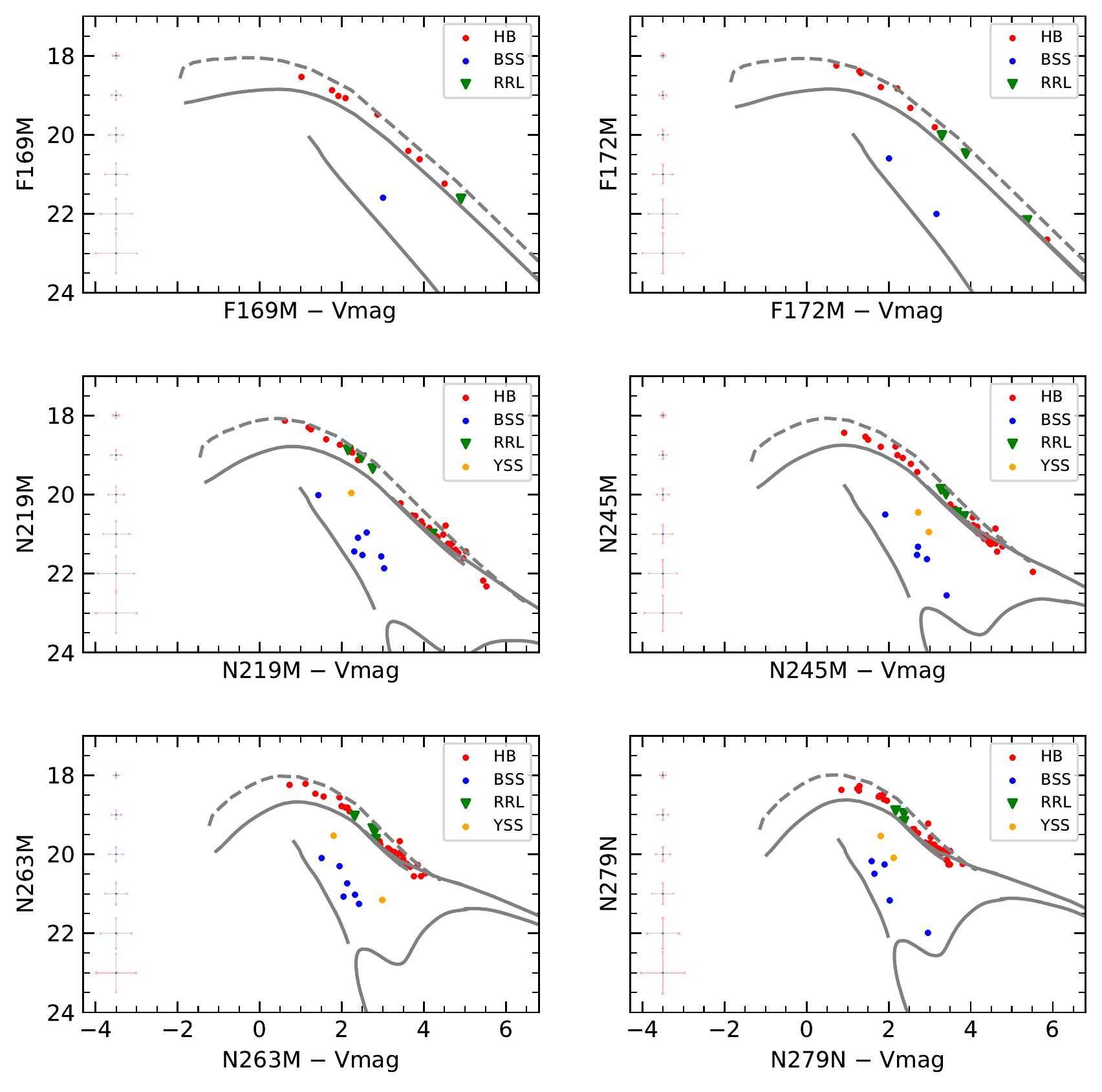}
     \caption{UV-optical CMDs of NGC 1261 after cross-matching ground based photometric (\protect\citealp{2010A&A...516A..23K}) and Gaia data with UVIT data in 4 NUV and 2 FUV filters. Rest of the details are same as in figure~\protect\ref{optuvhstcmds}.
     }
    \label{optuvgaiacmds}
\end{figure*}

 In Figure~\ref{optuvhstcmds}, the top two panels display the FUV-optical CMDs, generated using the F169M and F172M filters. The overlaid isochrone helps in defining the location of HB stars and BSSs, which span a large range in colour and magnitude, when compared to the optical CMDs, suggesting FUV-optical CMDs have better resolution in colour at a given magnitude. The HB stars bluer and redder to the RR Lyrae stars are detected in the FUV, suggesting that the BHB and the RHB stars are detected. The HB sequence is well aligned with the isochrone suggesting that the predicted and observed FUV magnitudes match well. In the FUV, the BSSs are found to span a large range in magnitude, for a given colour, and vice versa. The rest of the 4 panels show the NUV-optical CMDs. In NUV-optical CMDs in Figure~\ref{optuvhstcmds}, we find that the BSSs are as hot as blue HB stars and also they span a wide range in colour and magnitude in all the NUV CMDs. We have detected full HB population in all NUV-optical CMDs. The HB population appears as a tight sequence more-or-less aligned with the isochrone. In the case of CMDs using N245M, N263M and N279N filters, the red end of the HB is fainter than the isochrone. As the photometric errors are also large at this limiting magnitude, our data is only suggestive. We also note that same stars are fainter than the isochrone in both CMDs where N263M and N279N magnitudes are shown. 
 In all the CMDs, we have detected two stars at the blue extreme end of the HB. These stars are quite separated from the observed HB sequence and are likely to be very hot HB stars, as suggested by their UV-optical colour.\\
 
 \begin{table}
	\centering
	\caption{Number of detected HB stars and BSSs in different UVIT filters. Here $N_{HB}$, $N_{BSS}$, $N_{RRL}$ and $N_{Sx Phe}$ represents number of detected HB, BSS, RRL and Sx Phe stars respectively. The number of stars detected in the outer region ($>$ 3.4$'$ diameter) are shown in parentheses. }
	\label{tab:2}
\begin{tabular}{cccccc} 
		\hline
		\hline
		 Filter & $N_{HB}$ & $N_{BSS}$  & $N_{RRL}$ & $N_{Sx Phe}$ \\
		\hline
		 F169M & 59(9) & 41(1) & 10(1) & 1 \\
		 F172M & 97(11) & 55(2) & 15(3) & 1 \\
		 N219M & 196(40) & 71(8) & 19(4) & 2 \\
		 N245M & 216(46) & 75(7) & 19(4) & 2 \\
	 	 N263M & 222(47) & 62(8) & 19(4) & 1 \\
		 N279N & 221(46) & 70(7) & 18(3) & 3 \\
		\hline
	\end{tabular}
\end{table}
 
 
 
 To identify stars in the outer region of the cluster, UVIT detected stars were first cross-matched with Gaia proper motion membership data provided by \cite{2019MNRAS.488.3024B}, followed by cross-match with the ground based photometric data (\citealp{2010A&A...516A..23K}). Here, the photometric system adopted to create optical CMD is standard Johnson-Cousin photometric system. To generate UV-optical CMDs, we have converted the vega magnitude system into the AB magnitude system using photometric calibration mentioned in \citealp {2007AJ....133..734B}. Our sample of HB and BSSs detected with UVIT is 90\% complete in the outer region, when compared to the number of stars detected with Gaia. The 10\% of the stars, which are not detected, are fainter than the detection limit of the UVIT. Figures~\ref{optgaiacmd} and \ref{optuvgaiacmds} present the optical and UV-optical CMDs, respectively. We can clearly notice that less number of HB stars and BSSs are detected in the outer region of the cluster (outside the $3.4\arcmin$ diameter), when compared to the inner region. More stars are detected in the NUV when compared to the FUV. We detect only BHB stars in FUV pass bands and both BHB and RHB stars in all the NUV pass bands. We have also detected yellow straggler stars (YSS), identified based on their location in the optical CMD (figure~\ref{optgaiacmd}). We also detect very less number of BSSs in outer region as compared to inner region of the cluster. We notice that the detected BSSs are all redder than the predicted BSS line, though this could be an artefact due to the less number of detected BSSs.
 The comparison of the number of HB stars and BSSs detected in inner and outer region suggests that these stars may be segregated towards the center of the cluster.

\section{Spectral Energy Distributions of bright HB stars}
\label{sec:4}
Two extreme HB stars are detected in the central region. As these stars are well separated from the rest of the HB, we aim to check the evolutionary status of these stars by estimating their stellar parameters. It is also important to compare the properties of the BHB with these extreme HB stars. In order to estimate the parameters like effective temperature ($T_{eff}$), luminosity ($\frac{L}{L_{\odot}}$) and radius ($\frac{R}{R_{\odot}}$) of the UV bright HB stars, we have constructed their spectral energy distributions (SEDs). SEDs are constructed with the observed photometric data points from FUV-to-Optical and fitted with selected theoretical models as discussed below. We have used the virtual observatory tool, VOSA (VO Sed Analyser, \citealp{2008A&A...492..277B}) for SED analysis. VOSA utilises the filter transmission curves to calculate the synthetic photometry of the selected theoretical model. By using the fixed distance to the cluster, synthetic fluxes are scaled with the observed fluxes. After constructing the synthetic SED, it performs a $\chi^2$ minimisation test to compare the observed with the synthetic photometry to find the best fit parameters of the SED. The expression used to estimate reduced $\chi^2_{red}$ is
\begin{equation*}
\hspace*{2.0cm}
   \chi^2_{red} = \frac{1}{N-N_{f}} \displaystyle \sum_{i=1}^{N} {\frac{(F_{o,i} - M_{d}F_{m,i})^2}{\sigma_{o,i}^2}}
\end{equation*}
where N is the number of photometric data points, $N_{f}$ is the number of free parameters in the model, $F_{o,i}$ is the observed flux, $M_{d}F_{m,i}$ is the model flux of the star,
$M_{d} = \big(\frac{R}{D}\big)^2$ is the scaling factor corresponding to the star (where R is the radius of the star and D is the distance to the star) and $\sigma_{o,i}$ is the error in the observed flux. The number of photometric data points N for stars varies from 9 to 13 depending upon their detection in different available filters. The number of free parameters ($N_{f}$) used to fit SED are \big[Fe/H\big], log(g) and effective temperature. The radius of the stars were calculated by using scaling factor, $M_{d}$. As discussed below, \big[Fe/H\big] and log(g) are fixed and are not anymore free parameters in fitting the SED. \\

\begin{figure*}
\centering
\begin{subfigure}{0.5\textwidth}
  \centering
  \includegraphics[height=8.0cm,width=9.0cm]{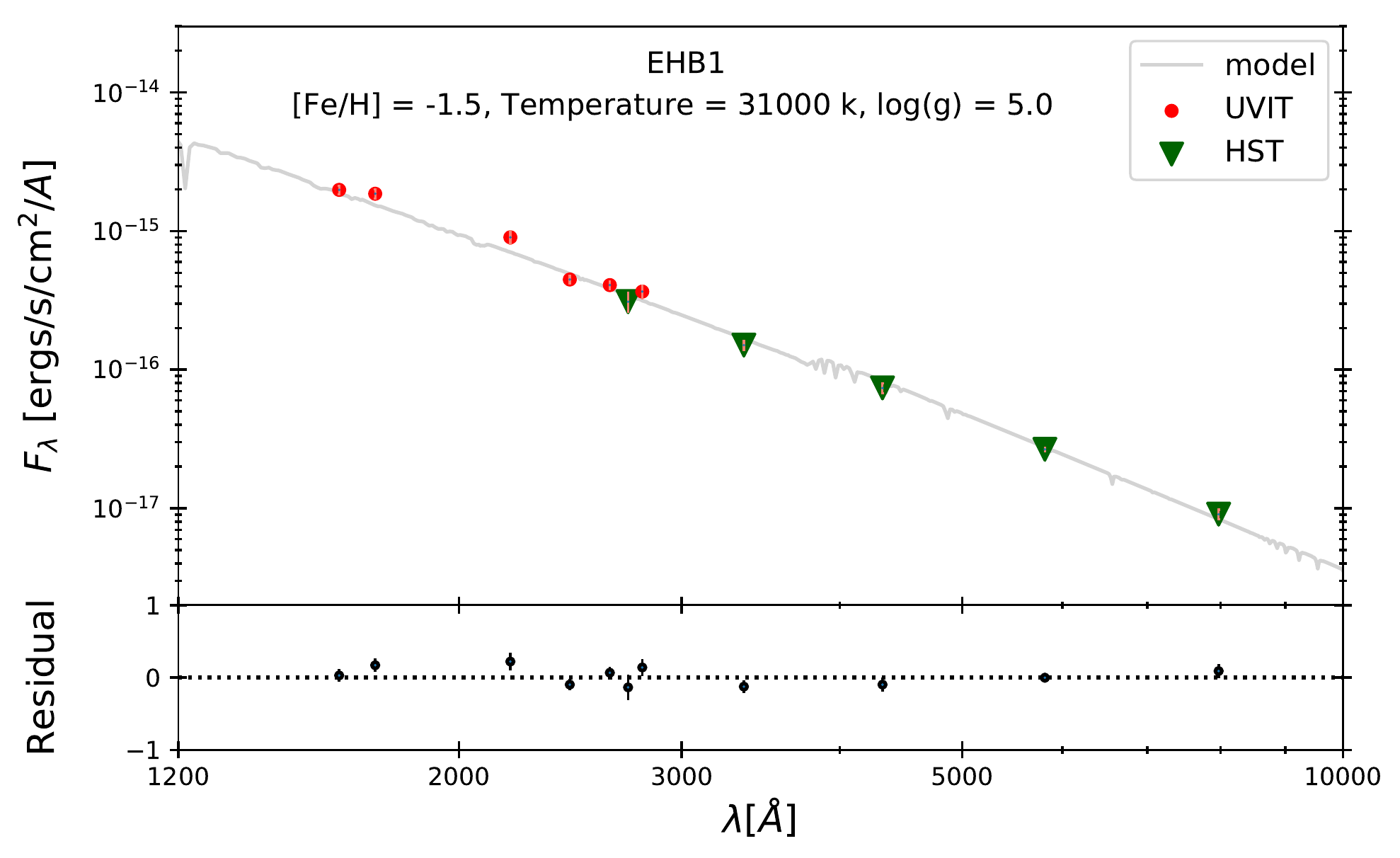}
  \end{subfigure}%
\begin{subfigure}{0.5\textwidth}
  \centering
  \includegraphics[height=8.0cm,width=9.0cm]{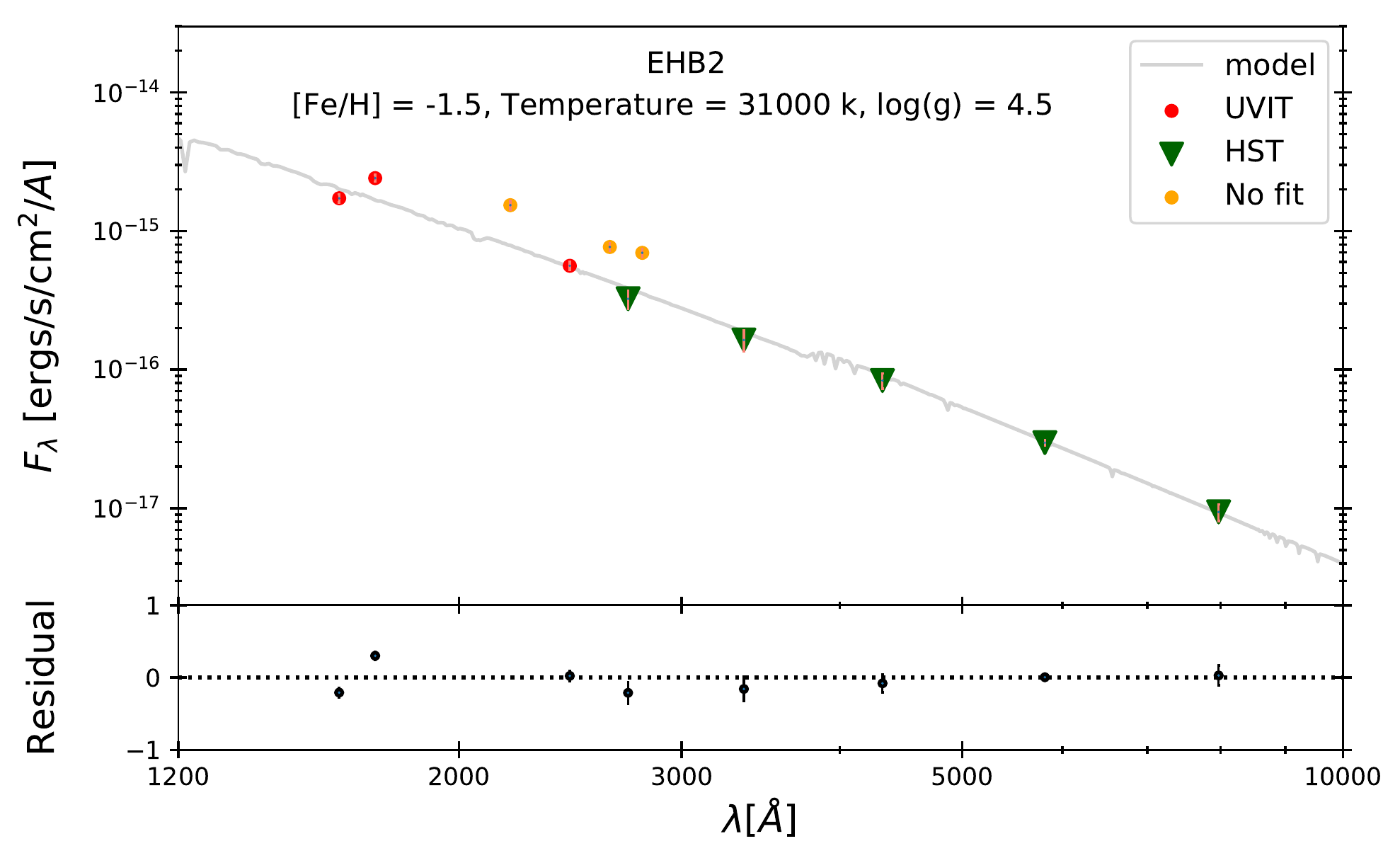}
\end{subfigure}
\caption{Spectral energy distribution (SED) of EHB1 and EHB2 stars after correcting for extinction. The best fit parameters are shown in the figure.}
\label{seds}
\end{figure*}

We have used Kurucz stellar atmospheric models to generate SEDs \citep{1997A&A...318..841C, 2003IAUS..210P.A20C} for the bright HBs which cover the UV to IR wavelength range. We fixed the value of metallicity \big[Fe/H\big] = $-$1.5, close to the cluster metallicity and had given the range of effective temperature from 5000-50000 K and corresponding log $g$ from 3-5 dex for the adopted Kurucz models to fit the SED of HBs (see Figure 19 in \citealp{2001MNRAS.324..937P}). 
We have combined 6 UVIT photometric data points  with 5 HST photometric data points from \cite{2018MNRAS.481.3382N} to generate SED for UV bright HB stars detected in the inner region of the cluster. For those detected in the outer region of the cluster, the photometric data points of UVIT (6 pass bands) with GAIA (3 pass bands) (\citealp{2018A&A...616A..12G}) and ground photometry (4 pass bands) (\citealp{2010A&A...516A..23K}) were combined. As errors in the HST measured flux are small which blows up the $\chi^2$ value for the HST data points, we assumed no error, which does not impact the fit parameters. In case, there are photometric data points with a zero observational error, VOSA assumes the biggest relative error present in the SED. VOSA makes use of Fitzpatrick reddening relation \citep{1999PASP..111...63F, 2005ApJ...619..931I} to correct for extinction in observed data points. The extinction corrected VOSA magnitudes are in a good agreement with that described in Section~\ref{sec:3}.\\

Best SED fits were obtained for 26 HB stars, out of which 2 are EHB stars and 24 are BHB stars. The SEDs of 2 EHB stars are shown in Figure~\ref{seds}. The estimated values of parameters  $T_{eff}$, $\frac{R}{R_{\odot}}$ and $\frac{L}{L_{\odot}}$ corresponding to the best fitting Kurucz model spectrum along with the errors for 26 HB stars are shown in Table~\ref{tab:3}. 
In left panel of Figure~\ref{seds}, we notice that the observed data points are well fitted with the Kurucz model spectrum, but in the case of EHB2 star, three UVIT data points are not fitted with the model spectrum. There seems to be some amount of excess flux in these UVIT filters, as also reflected in figure \ref{optuvhstcmds}. We excluded these three data points to make the good fit of the observed data points to the Kurucz model spectrum. The high temperatures of EHB1 (31000 K) and EHB2 (31000 K) suggest that they belong to the class of EHB stars (\citealp{1986A&A...155...33H}). The BHB stars have a range of temperature, 8000 - 12750 K, with one star with 12750 K. The SEDs of these stars are presented in Appendix-A.\\

\begin{figure}
    \hspace{-0.8cm}
    \includegraphics[height=9.5cm,width=9.5cm]{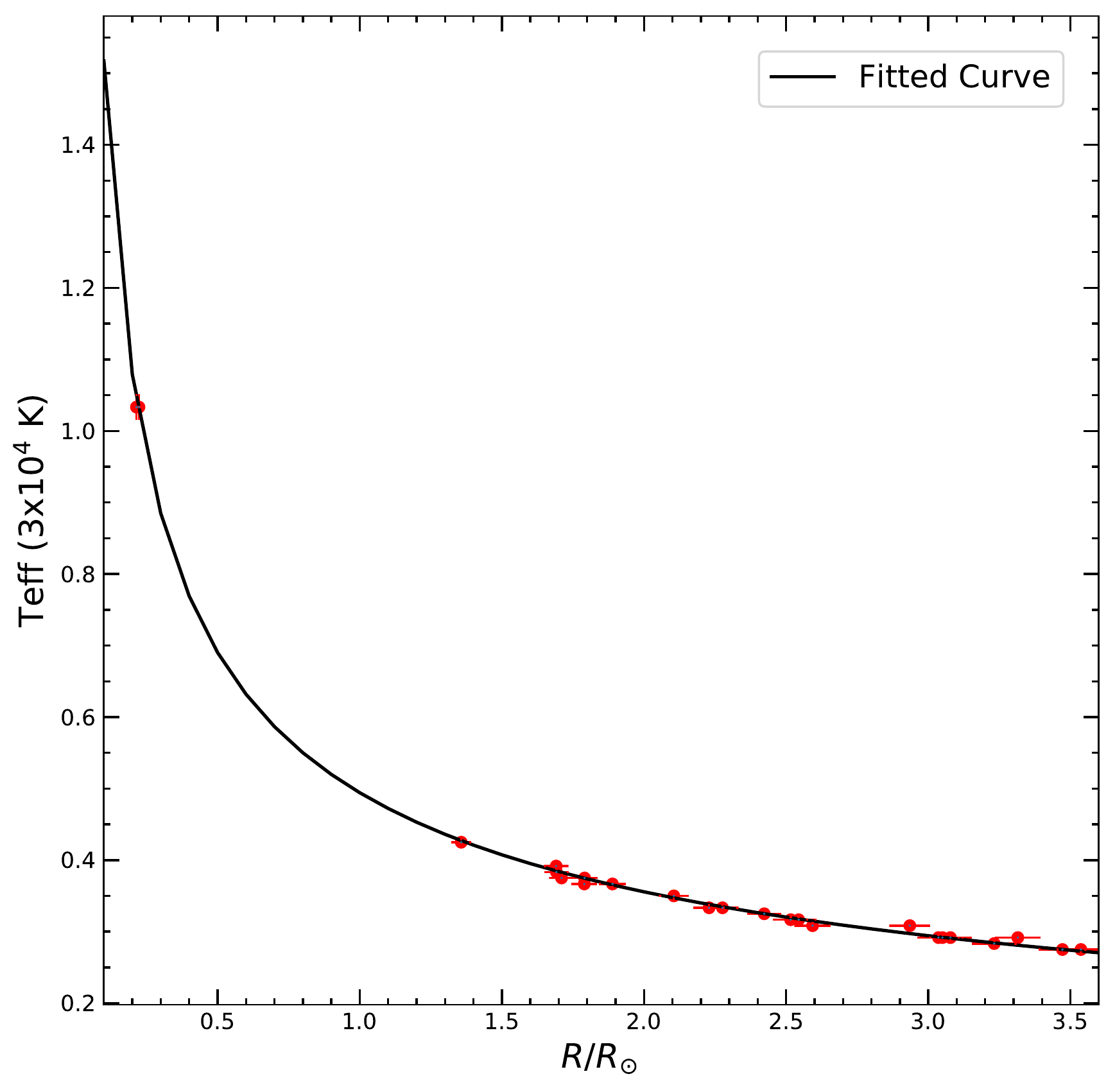}
    \caption{ Variation of radius of all bright HB stars with their effective temperatures, determined from SEDs. Red filled circles are representing bright HB stars and the black curve representing the function $T_{eff} = \frac{0.5}{\sqrt{R}} + 0.002$, is the fitted curve to the observed distribution.}
    \label{tempradvar}
\end{figure}

We also note from Table~\ref{tab:3} that the luminosity, $\frac{L}{L_{\odot}}$, is more or less constant for the HB stars (except for a few cases), whereas the $T_{eff}$ is found to increase with a corresponding decrease in radius. 
As we have a good number of BHB stars, we estimate this relation assuming a constant luminosity. 
The Figure~\ref{tempradvar} shows the variation of effective temperature of all bright HB stars with their corresponding radius. The best-fit relation to the observed points plotted in Figure~\ref{tempradvar} is :
\begin{equation*}
\hspace{2cm}
    T_{eff} = 0.5\sqrt{\frac{R}{R_{\odot}}} + 0.002
\end{equation*}
This relation corresponds to the well-known Stefan-Boltzmann law. Since the intrinsic luminosity of the BHB and EHB stars is not very different, so they should satisfy a quite similar $T_{eff}$-R relation.
We therefore extended the plot to higher $T_{eff}$ and included the EHB stars. 
This relation is found to fit the EHB stars quite well as shown in the black curve in Figure~\ref{tempradvar}. The value of goodness of fit parameter for this curve-fit is 0.99, which is close to 1, hence, a very good fit. This fit also confirms the accuracy of the derived parameters from the SED fit.\\

\begin{figure}
\hspace{-0.8cm}
\includegraphics[height=9.5cm,width=9.5cm]{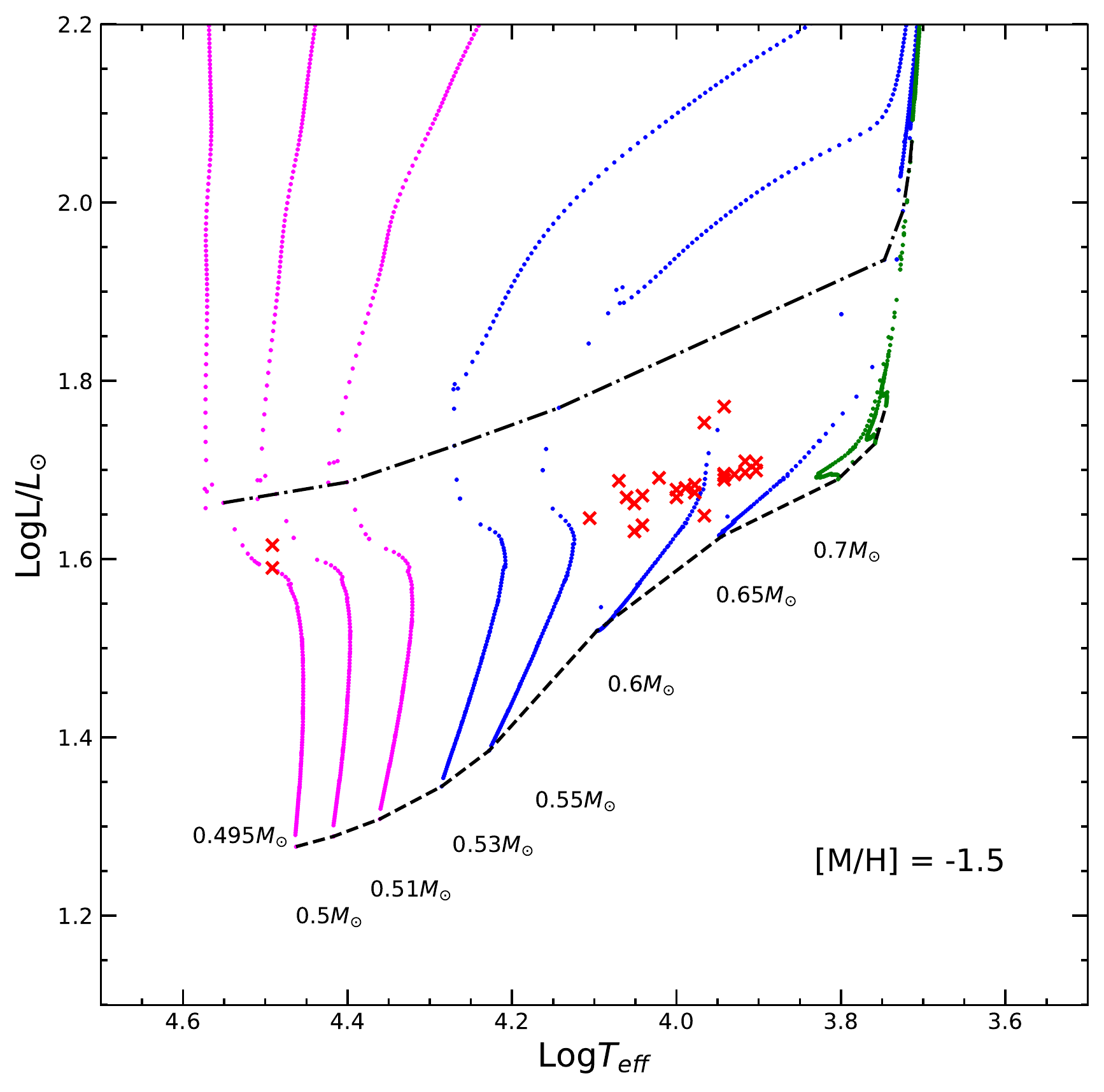}
\caption{Evolutionary tracks corresponding to different masses with metallicity close to cluster metallicity. Evolution starting from HB phase to the moment when star has entered to the post-HB phase (\protect\citealp{2019A&A...627A..34M}) are shown. In the plot, magenta, blue and green colours represent the sequences populating the extreme, blue and red parts of the HB. Black dashed and dash-dotted line indicates the ZAHB and TAHB, respectively. Bright HB stars identified with UVIT are shown with red cross symbols.}
\label{evotrack}
\end{figure}

In order to check the evolutionary status of EHB stars identified with UVIT, we have plotted the theoretical evolutionary tracks using the models presented by \cite{2019A&A...627A..34M}. The models used in their paper were computed by the extension of the PAGB evolutionary models by \cite{2016A&A...588A..25M}. We have selected the model with metallicity close to the cluster metallicity. The theoretical evolutionary tracks corresponding to different masses starting from the zero age horizontal branch (ZAHB) through to a point late in post-HB evolution or a point on the PAGB cooling track are shown in Figure~\ref{evotrack}. The terminal age HB (TAHB) representing the end of the HB phase is shown with dash-dotted line in Figure~\ref{evotrack}. We can see in Figure~\ref{evotrack} that all BHB stars are lying along the BHB tracks. Figure~\ref{evotrack} also shows that two EHB stars are indeed found along EHB tracks lying below the TAHB. So, these two stars do not belong to any of the  evolved class of stars, which lie on hot PAGB cooling sequence. This suggests that the two EHB stars share the properties of the BHB stars and therefore can be considered as the extreme extension of the HB. They are therefore still in the HB evolutionary phase. From the model, the masses of these two stars turn out to be approximately 0.5M$_{\odot}$ corresponding to the core mass of the stars lying on the HB.

\begin{table*}
\centering
     \caption{SED fit parameters of bright HB stars detected using UVIT data.}
    \label{tab:3}
    \large
	\makebox[0.79\linewidth]
	{
	\begin{tabular}{cccccccc} 
		\hline
		\hline
		 Star ID & RA (deg) & DEC (deg) & $T_{eff}$ (K) & $\frac{L}{L_{\odot}}$  & $\frac{R}{R_{\odot}}$ & ${\chi}_{red}^2$ \\
		\hline
		 EHB1 & 48.04181 & -55.20597 & $31000 \pm 500$ & $38.92 \pm 1.09$ & $0.21 \pm 0.01$  & 2.7\\
		 EHB2 & 48.07732 & -55.22445 & $31000 \pm 500$ & $41.28 \pm 0.9$ & $0.22 \pm 0.01$ & 8.6 \\
		 BHB1 & 48.07928 & -55.22457 & $12750 \pm 125$ &  $44.26 \pm 1.64$ & $1.36 \pm 0.03$ & 4.6 \\
		 BHB2 & 48.09452 & -55.2267 & $11000 \pm 125$ & $43.46 \pm 1.96$ & $1.79 \pm 0.04$ & 6.9 \\
	 	 BHB3 & 48.08742 & -55.22956 & $11750 \pm 125$ &  $48.75 \pm 2.58$ & $1.69\pm 0.04$ & 2.9  \\
		 BHB4 & 48.06405 & -55.21413 & $11250 \pm 125$ & $42.77 \pm 1.87$  & $1.71 \pm 0.04$ & 3.6 \\
		 BHB5 & 48.07904 & -55.21753 & $11250 \pm 125$ & $45.97 \pm 2.25$ & $1.79 \pm 0.04$ & 4.3\\
		 BHB6 & 48.05424 & -55.21919 & $11000 \pm 125$ & $46.91 \pm 1.87$ & $1.89 \pm 0.04$ & 4.5 \\
		 BHB7 & 48.05896 & -55.21738 & $10500 \pm 125$ &  $49.12 \pm 2.51$ & $2.11 \pm 0.05$ & 7.3 \\
		 BHB8 & 48.08082 & -55.22729 & $10000 \pm 125$ & $46.70 \pm 2.09$ & $2.23 \pm 0.05$ & 6.3\\
		 BHB9 & 48.06626 & -55.21944 & $9500 \pm 125$ & $47.28 \pm 2.06$ & $2.52 \pm 0.06$ & 6.8\\
		 BHB10 & 48.06391 & -55.21627 & $9250 \pm 125$ &  $44.54 \pm 1.59$ & $2.59 \pm 0.06$ & 5.9\\
		 BHB11 & 48.04864 & -55.21555 & $9250 \pm 125$ & $56.64 \pm 1.89$ & $2.94 \pm 0.07$ & 4.2 \\
		 BHB12 & 48.05479 & -55.22369 & $8750 \pm 125$ & $48.88 \pm 1.41$ & $3.05 \pm 0.07$ & 3.6 \\
		 BHB13 & 48.10376 & -55.21723 & $8500 \pm 125$ & $49.55 \pm 1.57$ & $3.23 \pm 0.08$ & 5.0\\
		 BHB14 & 48.08487 & -55.22036 & $8750 \pm 125$ & $49.63 \pm 1.83$ & $3.08 \pm 0.07$ & 2.1\\
		 BHB15 & 48.06478 & -55.21766 & $8750 \pm 125$ & $49.42 \pm 1.64$ & $3.04 \pm 0.07$ & 5.5\\
		 BHB16 & 48.07389 & -55.22981 & $8250 \pm 125$ &  $49.79 \pm 1.69$& $3.47 \pm 0.08$ & 4.1\\
		 BHB17 & 48.04274 & -55.22656 & $8000 \pm 125$ &  $51.07 \pm 1.85$ & $3.73 \pm 0.09$ & 3.4 \\
		 BHB18 & 48.06466 & -55.21258 & $8000 \pm 125$ &  $50.06 \pm 1.55$ & $3.69 \pm 0.09$ & 5.0\\
		 BHB19 & 48.05484 & -55.18698 & $11500 \pm 125$ &  $46.69 \pm 3.98$ & $1.69 \pm 0.04$ & 5.7\\
		 BHB20 & 48.14102 & -55.20569 & $10000 \pm 125$ & $47.64 \pm 4.98$& $2.28 \pm 0.05$ & 4.6\\
		 BHB21 & 48.01035 & -55.31072 & $9750 \pm 125$ & $47.85 \pm 5.55$ & $2.42 \pm 0.06$ & 3.9 \\
		 BHB22 & 48.12378 & -55.25798 & $9500 \pm 125$ & $48.24 \pm 7.88$ & $2.54 \pm 0.06$ & 1.8 \\
		 BHB23 & 47.90097 & -55.23117 & $8750 \pm 125$ & $59.04 \pm 10.44$ & $3.32 \pm 0.07$ & 1.5\\
		 BHB24 & 48.14307 & -55.27254 & $8250 \pm 125$ & $51.27 \pm 13.41$ & $3.54 \pm 0.08$ & 1.8 \\
		\hline
	\end{tabular}
	}
\end{table*}


\section{Discussion}
\label{sec:5}
In this study, we used data from UVIT onboard Astrosat satellite to create NUV and FUV CMDs and to investigate the morphology of HB stars. UVIT is not able to detect stars in the MS, RGB and SGB evolutionary phases, since these stars are fainter than 22 mag in UV. To generate UV CMDs, we cross-matched UVIT detected stars with HST data for inner region corresponding to $\sim3.4'$ diameter and with ground based data along with GAIA for outer region as UVIT has a large circular field of view of $28{\arcmin}$ diameter. The optical and UV CMDs are created for the proper motion members of the cluster. In all UV-optical CMDs, we find that the HB stars do not show the horizontal distribution, as found in the  optical CMDs. This is due to the fact that the HB stars have a range in temperature and the flux in UV pass bands are more sensitive to temperature than in optical. The HB feature is therefore found to be slanting and is similar to that found in the GALEX UV CMDs obtained by \cite{2012AJ....143..121S}. They suggest that the slope of HB is mainly a result of bolometric correction effects. \citealp{2013MNRAS.430..459D} investigated the HB morphology of three GCs, namely, M3, M13 and M79, using HST data in optical and UV bandpasses, and they also found the similar HB distribution in Far-UV CMDs (see their Fig. 2). The UV magnitude distribution of the HB stars is found to closely match with the over-plotted isochrone, within error, in all UV-optical CMDs. As we detect the full stretch of the HB including the RHB, RRL and BHB, we find a good match for the UV magnitudes between the observation and the updated BaSTI-IAC models.\\

A large gap is observed between EHBs and BHBs, and that needs to be explored. However, in massive and very dense GCs such as NGC 2808, NGC 6752, $\omega$ cen,  a large number of EHB stars are found with a well defined sequence in the optical CMDs. In less massive clusters, they are less in number and are hard to detect at the faint end of the HB in optical CMDs. In these cases, UV CMDs play an important role since EHB stars are bright in UV CMDs and also follow a separate sequence. The clusters with a few stars bluer than the M-jump are NGC 5466, NGC 6981, NGC 3201, NGC 2298 etc. (see Figure 5 in \citealp{2016ApJ...822...44B}). We have detected full HB in NUV and only hot HB stars in FUV. Two EHB stars are also identified in all UV-optical CMDs, which have bluer colours as compared to other hot HB stars. \cite{2015MNRAS.451..275V} studied 48 galactic GCs to explore the radial distribution of multiple stellar populations known to exist in clusters. They identified 6 EHB stars in NGC 1261, which may or may not be the members of the cluster. They also used the different selection criteria to select different HB sub-populations. The HB stars identified by UVIT are proper motion members of the cluster, including the two EHB stars.\\

The bright HB stars are found to have a range of temperature, from 8000 K to 12750 K, with the hot end very close to G-jump (\citealp{1999ApJ...524..242G}) ($T_{eff}$ = 11,500 K), in the HB distribution . One star is found to have 12750 K along with the two hotter stars. The hot end of the HB distribution therefore, coincides with the G-jump, in this cluster. There is only one star slightly hotter and there is no further extension of the HB to the EHB, beyond the G-jump. Therefore, we do not detect M-jump (\citealp{2002ApJ...576L..65M, 2004A&A...420..605M}), expected at $\sim$ 23,000 K. Instead, we detect two stars with $T_{eff}$ = 31,000 K. These coincide with the $T_{eff}$ expected for the gap between EHB and the blue-hook stars. 
The cluster NGC 1261, therefore has an HB distribution truncating at the G-jump at the hot end, and with the presence of two EHB stars.\\

With the estimation of the surface parameters of the BHB stars, we are able to derive a relation between the temperature and radius, as the bolometric luminosity is found to be almost the same for these stars. We find that the temperature and radii of the two EHB stars agree with the relation derived above, suggesting that they can be considered as the extreme end of the HB population.  Thus, in this cluster the EHB stars are normal core He burning stars like other HB stars and are not in any different evolutionary phase. This is supported by their position in the HR diagram overlaid with isochrones. It must be noted that these two EHB stars may be starting to evolve off from the HB.\\

The canonical picture of the EHB stars is established by \cite{1986A&A...155...33H}, in which EHB stars are helium core burning stars with masses close to the core helium flash mass of 0.47M$_{\odot}$, and an extremely thin hydrogen envelope, not more than, 1$\%$ by mass. Our estimation of mass for the detected EHB stars match well with this definition. Although the evolution of EHB stars after the exhaustion of Helium in the core is better understood, the formation pathways that lead to the EHB is much less understood. Many have explored the binary nature of EHB stars and a wide variety of companions are found as the binary fraction among the EHB stars are found to be much higher than normal stars. The way in which the envelope is lost, is mostly attributed to binary interactions, depending on the mass ratio of the system. \cite{2015MNRAS.449.2741L} suggested that tidally enhanced stellar wind in binary evolution is able to provide the enhanced mass loss on the RGB needed for the late hot flash scenario to explain the formation of blue-hook stars. They have adopted different initial orbital periods for binaries to explain the formation of canonical HB and blue-hook stars. The EHB stars detected in this cluster are likely to be single stars, as suggested by the SEDs. The SEDs are well fitted by a single spectrum with minimum residual across the wavelength. It is proposed by \cite{1993ApJ...407..649C} that the delayed helium flash (HEF) is a promosing scenario to explain the existence of EHB stars. Due to high mass loss during RGB evolutionary phase, a star will lose so much envelope mass that it will not be able to ignite the helium flash at the tip of the RGB, thus evolving towards the WD cooling sequence with an electron degenerate core. Depending on the residual mass of the envelope, a star will undergo He flash either at the bright end of the WD cooling sequence, known as early hot flasher (EHF) or along the WD cooling sequence, known as late hot flasher (LHF). After the He flash, these stars will settle on a blue-hook at the hot end of the HB. As these stars have hugely reduced envelope mass, they will be much hotter than their counterparts on the canonical ZAHB. \cite{1996ApJ...466..359D} also suggested hot He-flash scenario to explain the origin of EHB stars in globular clusters. \cite{2003ApJ...582L..43C} computed the models of the population II low mass stars through the Helium flash mixing (HEFM) phase, suggested that the HEFM scenario as a plausible explanation for the existence of blue hook stars.
\\

Here we summarise the possible scenarios for the formation of single EHB stars. 
\cite{1984ApJ...277..355W} and \cite{2000MNRAS.313..671S} suggested the merger of two He-core WDs as a possible mechanism. Another possibility is the trigger of CE ejection by a giant planet, that evaporates in the process (\citealp{1998AJ....116.1308S}). The fraction of close binaries among the EHB stars in GCs is found to be very small \citep{2006A&A...451..499M, 2009A&A...498..737M, 2011A&A...528A.127M}. It is possible that in old systems like the GCs, WD-WD merger may likely result in the formation of EHB stars, as a consequence they are single stars. If these two stars are the product of the merger of two He-core WDs, then they can be helium rich, which will lead to higher effective temperature compared to estimated using Kurucz stellar atmospheric models (\citealp{1984ApJ...278..702S}). Another model suggested for the EHB star formation is primordial enrichment in helium \citep{2002A&A...395...69D}. In this scenario, the EHB stars are produced via the normal evolution of He-enriched sub-populations in GCs. These sub-populations might have formed out of the material polluted by the ejecta from massive AGB stars. For a given age and metallicity, He enhanced stars have smaller masses than He normal stars, resulting in a bluer HB morphology. A super-solar surface He abundance cause a huge mass loss on RGB phase by increasing the RGB tip luminosity, but the phenomena responsible for a huge He enhancement invoke non-canonical mixing during the RGB stage, and dredge-up induced by H-shell instabilities \citep{1979ApJ...229..624S, 1988ApJ...324..840V, 1997ApJ...474L..23S, 2003ApJ...598.1246D}. The discovery of multiple stellar populations in GCs opened a new frontier, as one of the causes is the variation in Helium enrichment, with the blue HB stars produced by Helium rich stars \citep{2004MSAIS...5..105B,2005ApJ...621..777P,2007ApJ...661L..53P,2012ApJ...760...39P,2015AJ....149...91P}. The models of multiple populations with different helium abundances successfully reproduce both the MS splitting and the multi-modal HB morphology of both $\omega$ Cen (\citealp{2005ApJ...621L..57L}) and NGC 2808 (\citealp{2005ApJ...631..868D}). The He-enhancement thus represents a promising model, and alternative to the binary scenario, for the formation of EHB stars in GCs. We find that both the EHB stars are single stars and estimate their masses to be 0.495 M$_\odot$, which is quite close to the theoretical estimate for EHB stars. The possible formation mechanism for these stars is likely to be the enhanced mass loss in the RGB phase, either due to rotation or enhanced helium (primordial or mixing).\\

 The comprehensive studies of EHB and blue-hook stars by \cite{2001ApJ...562..368B,2010ApJ...718.1332B} suggest that these stars could be produced by delayed helium-flash after the RGB, and are termed as early hot-flashers and late hot-flashers. In the case of early hot-flashers, a star ignites helium during the evolution to the top of the WD cooling curve. Flash mixing will not happen between the envelope and the core of a star as the high entropy of a strong hydrogen-burning shell acts as a barrier that prevents the flash convection from penetrating into the envelope. Hence, the envelope mass or composition of a star will remain unchanged, which leads to more or less the same luminosity compared to the canonical HB stars ( See Figures 1 \& 4 in \citealp{2010ApJ...718.1332B}). Late hot-flash scenario results in the flash mixing between the hydrogen envelope and the helium core, which increases the helium and carbon abundances in the envelope. The flash mixing lowers the luminosity of a star with respect to the canonical EHB star. The EHB stars of this cluster are not sub-luminous with respect to the BHB stars and hence are unlikely to be late hot-flashers. These stars therefore could be early hot-flash objects, which follow the properties of the canonical HB stars. 
 Spectroscopic observations are required to explore the nature of these two stars and validate the formation scenarios described above. Since there is not enough mass for the hydrogen shell surrounding the core in EHB stars, these stars are likely to evolve directly to the white dwarf phase.\\

All the detected BSSs occupy a region parallel to HB sequence as well as span a wide range in both colour and magnitude, in the UV-optical CMDs. In general, the temperature  of BSSs ranges from 6000-8000K, but in this cluster, we notice that some of the BSSs have colours similar to the BHB stars, which in turn imply that BSSs may be as hot as the BHB stars. We detected a few BSSs in the outer region of the cluster beyond half-mass radius (r>r$_{h}$). These BSSs might have formed through mass-transfer in close binary systems, which dominates in a low-density environment. Mass-transfer models are required to confirm this. 
We also detect two YSSs in the outer part of the cluster. We plan to construct the SEDs of the BSSs as well as the YSSs to understand the properties of these stars.
\section{Summary and Conclusions}
\label{sec:6}
 In this paper, we present the photometric results of NGC 1261 imaged using the UVIT on ASTROSAT. The advantage of using UVIT over HST  is its large field of view covering the full cluster region and over GALEX is its good spatial resolution and multiple filters.
 We have characterized the HB member stars for the first time in this cluster using UVIT, HST, ground-based and GAIA data. Below, we summarise the important results from this study:
\begin{itemize}
    \item We constructed Optical and UV-optical CMDs of the member stars and overlaid with isochrones generated for respective filters. We detected only BHB and 2 EHB stars in the FUV CMDs whereas the full HB is detected in NUV CMDs. We also detected BSSs which span a wide range in magnitude as well as in colour in NUV CMDs.\\
    \item The effective temperatures, luminosities and radii of 24 BHB and 2 EHB stars are estimated by generating SED using multi-wavelength data. The $T_{eff}$ of BHB stars range from 8000-12750 K whereas EHB stars have $T_{eff}$ more than 30000 K. \\
    \item Keeping the L constant, we fitted a R versus $T_{eff}$ relation for the BHB stars, which is found to fit the EHB stars as well. Their location in HR diagram overlaid with isochrones confirm that they are EHB stars with a mass of $\sim$ 0.5M$_\odot$.\\

\end{itemize}
   Based on the results from the UVIT study of this cluster, we conclude that:
\begin{itemize}
    \item Most of the RHB stars are too faint to be detected by UVIT in FUV filters. HB stars form a tight sequence in the UV-optical CMDs which is largely fitted by the BaSTI-IAC isochrones.\\
    \item The EHB stars in this cluster are likely to be single stars. We constrain the formation pathways of these single EHB stars to extreme mass loss in the RGB phase (either due to rotation or enhanced Helium), OR early hot-flash scenario.\\ 
\end{itemize}
\section*{Acknowledgements}
We are thankful to the reviewer for the thoughtful comments and suggestions which improved the quality of the manuscript. UVIT project is a result of collaboration between IIA, Bengaluru, IUCAA, Pune, TIFR, Mumbai, several centres of ISRO, and CSA. This publication uses the data from the \textit{ASTROSAT} mission of the Indian Space Research  Organisation  (ISRO),  archived  at  the  Indian  Space  Science  Data Centre (ISSDC). We want to thank Iv\'an H. Bustos Fierro for providing us with GAIA proper motion membership data. This research made use of VOSA, developed under the Spanish Virtual Observatory project supported by the Spanish MINECO through grant AyA2017-84089. This research also made use of Topcat \citep{2005ASPC..347...29T, 2011ascl.soft01010T}, Matplotlib (\citealp{2007CSE.....9...90H}), NumPy (\citealp{2011CSE....13b..22V}), Scipy \citep{2007CSE.....9c..10O, article}, Astropy \citep{2018AJ....156..123A} and Pandas (\citealp{mckinney-proc-scipy-2010}).

\section*{Data availability}
The UVIT data used in this article will be shared on request.



\bibliographystyle{mnras}
\bibliography{references} 




\appendix

\section{SED Fitting for BHB Stars}
The procedure of SED fitting for bright HB stars is described in Section~\ref{sec:4}. The SEDs for 24 BHB stars are shown in Figure~\ref{sed1}. \\

\begin{figure*}
\centering
\begin{subfigure}{0.5\textwidth}
\centering
\includegraphics[height=7.5cm,width=9.0cm]{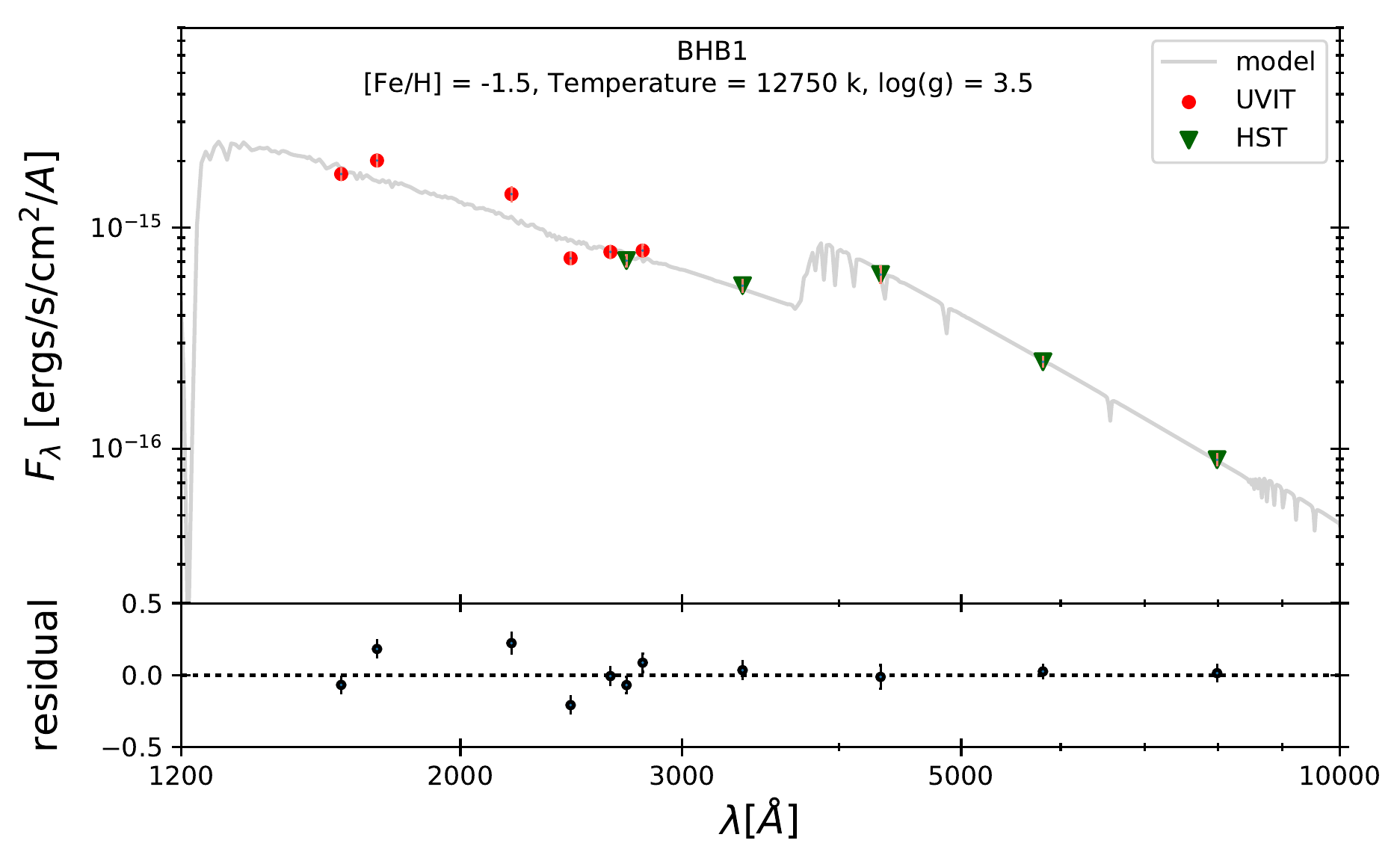}
\end{subfigure}%
\begin{subfigure}{0.5\textwidth}
\centering
\includegraphics[height=7.5cm,width=9.0cm]{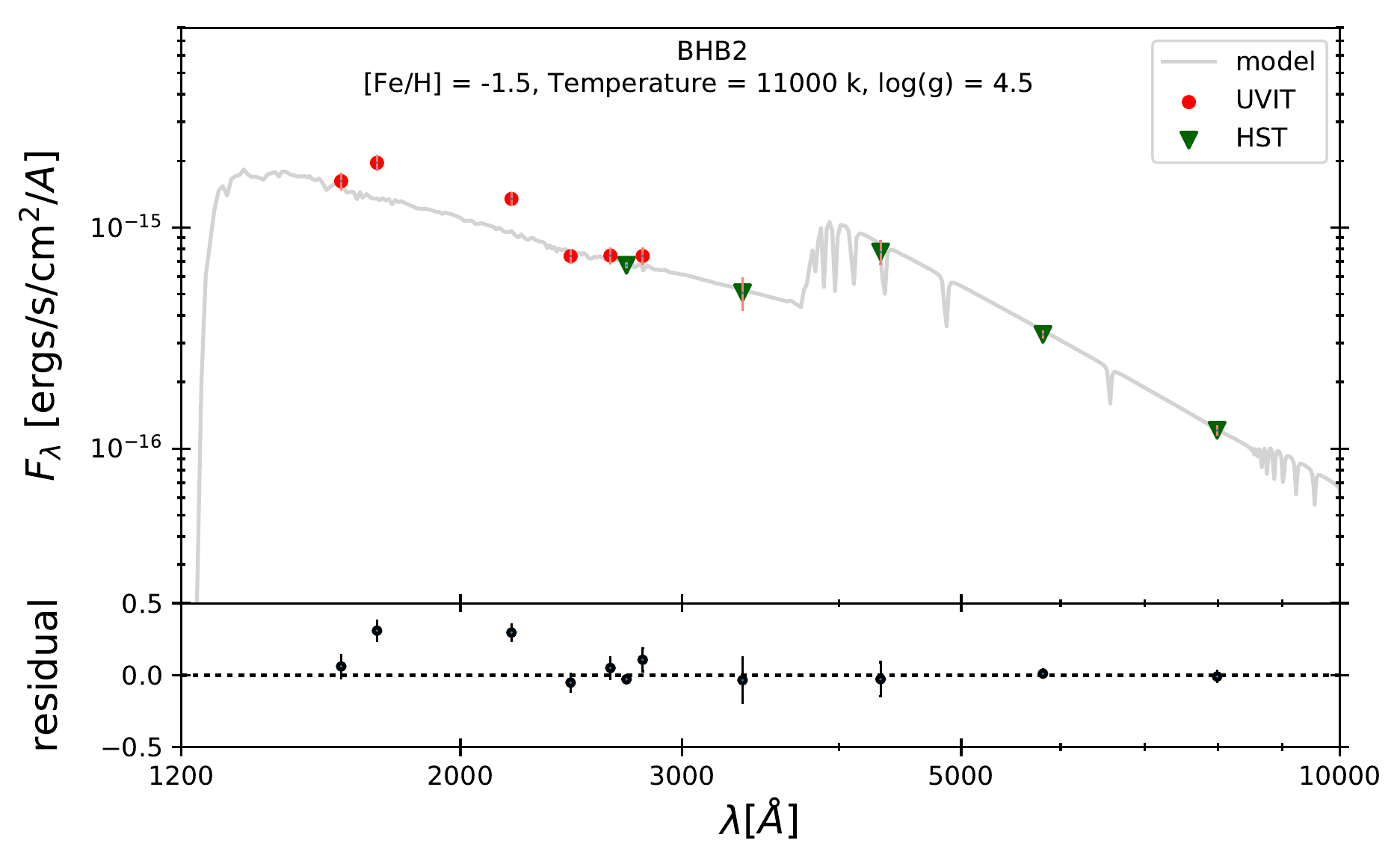}
\end{subfigure}
\begin{subfigure}{0.5\textwidth}
\centering
\includegraphics[height=7.5cm,width=9.0cm]{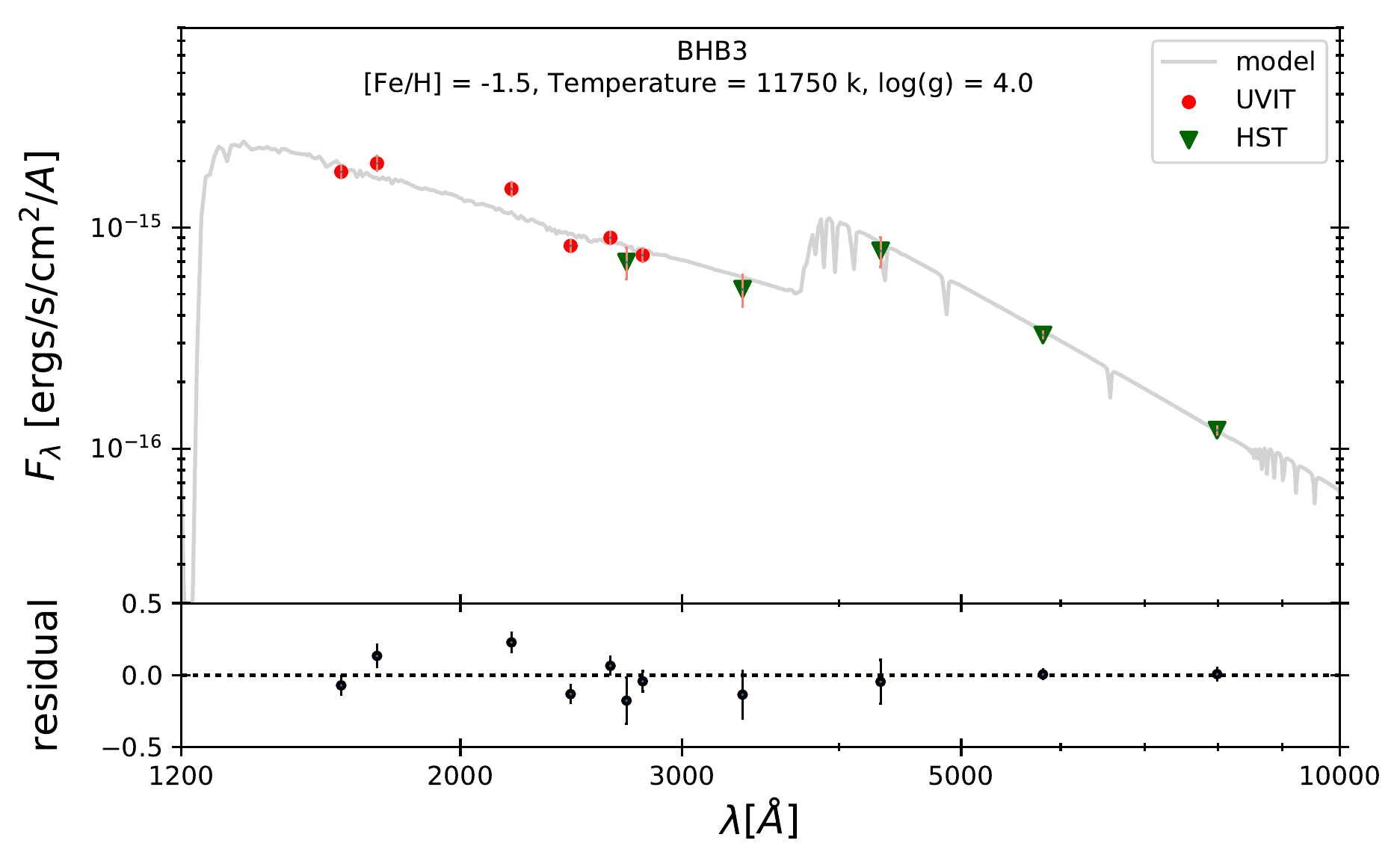}
\end{subfigure}%
\begin{subfigure}{0.5\textwidth}
\centering
\includegraphics[height=7.5cm,width=9.0cm]{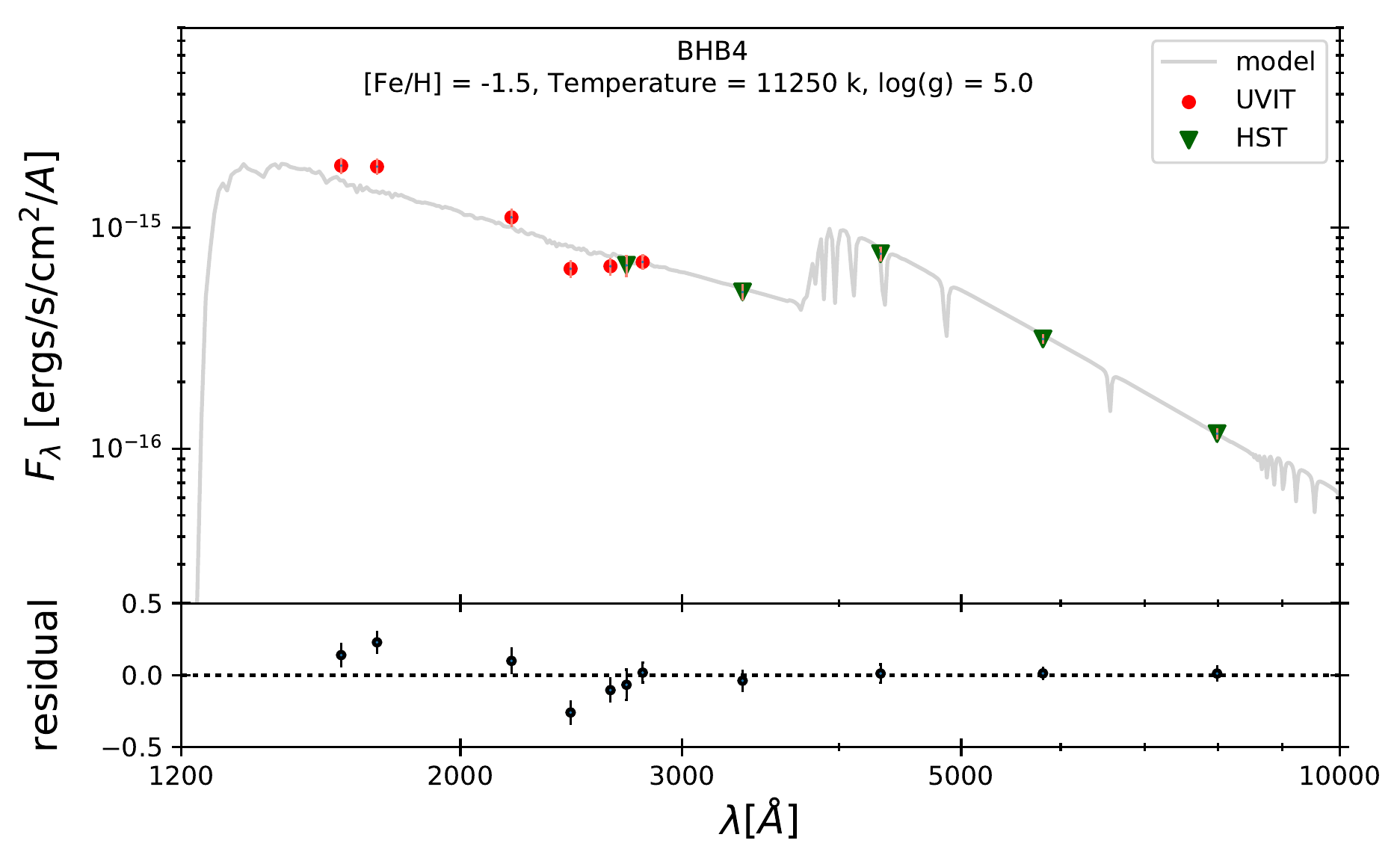}
\end{subfigure}
\begin{subfigure}{0.5\textwidth}
\centering
\includegraphics[height=7.5cm,width=9.0cm]{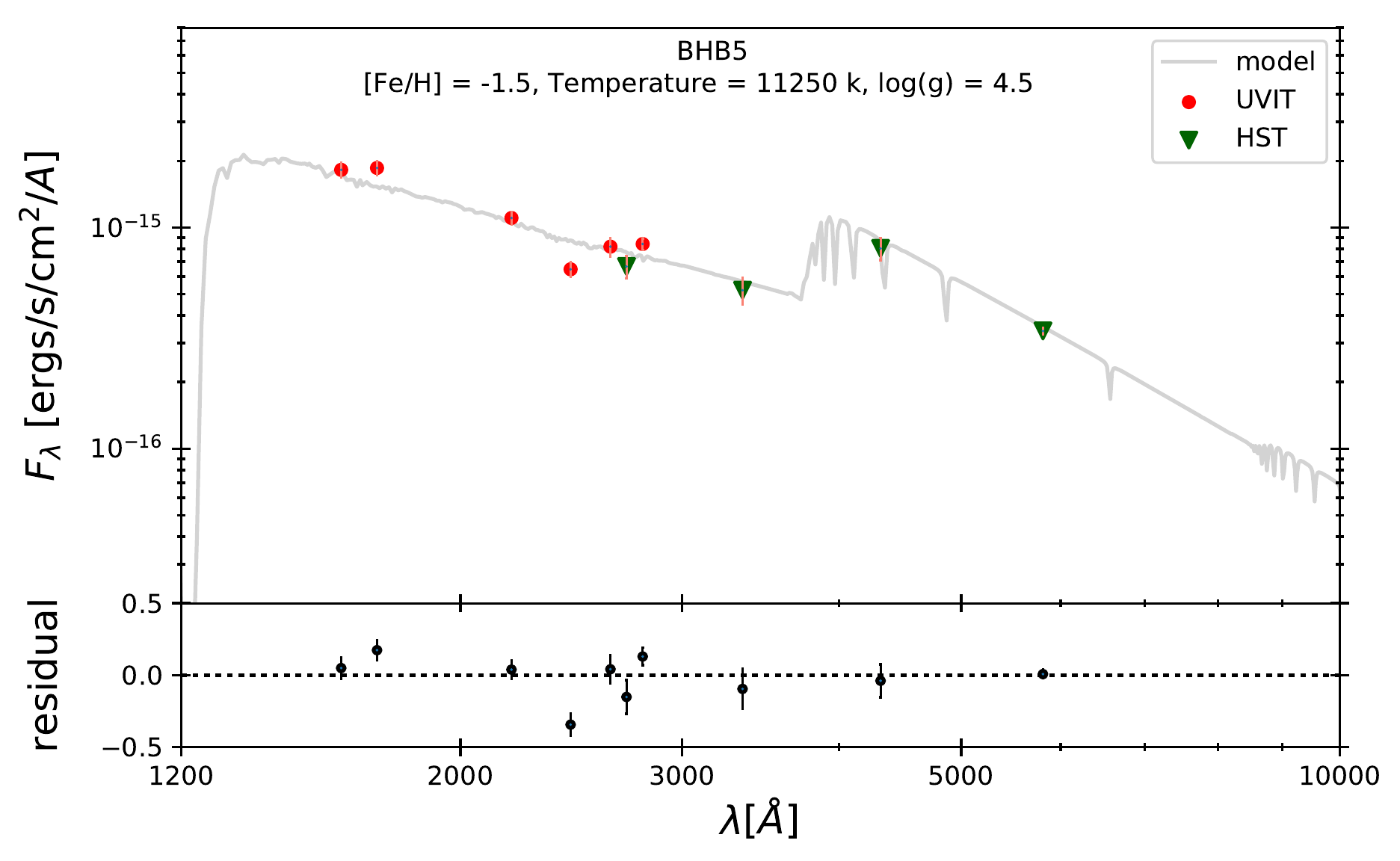}
\end{subfigure}%
\begin{subfigure}{0.5\textwidth}
\centering
\includegraphics[height=7.5cm,width=9.0cm]{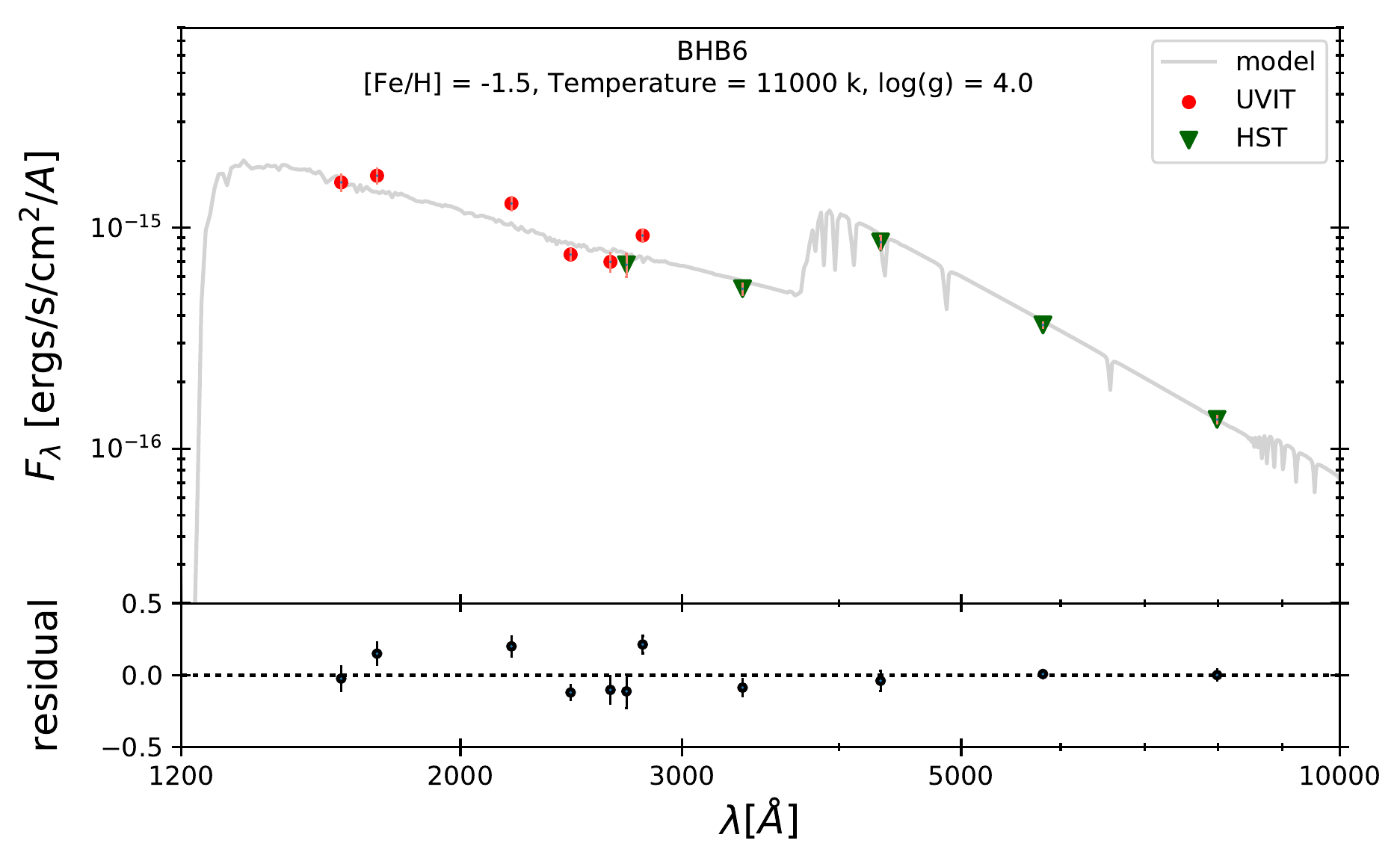}
\end{subfigure}%
\caption{Spectral energy distribution (SED) of all BHB stars after correcting for extinction. The best fit parameters are shown in the figure. The UVIT, HST data points used to create SEDs for stars lying in the region covered with HST are shown with red circles and green triangles respectively. For the stars lying in the outer region, UVIT, Ground based photometric and GAIA data points are shown with red circles, green triangles and cyan square respectively.}
\label{sed1}
\end{figure*}

\renewcommand{\thefigure}{A\arabic{figure}}
\addtocounter{figure}{-1}
\begin{figure*}
\centering
\begin{subfigure}{0.5\textwidth}
  \centering
  \includegraphics[height=8.0cm,width=9.0cm]{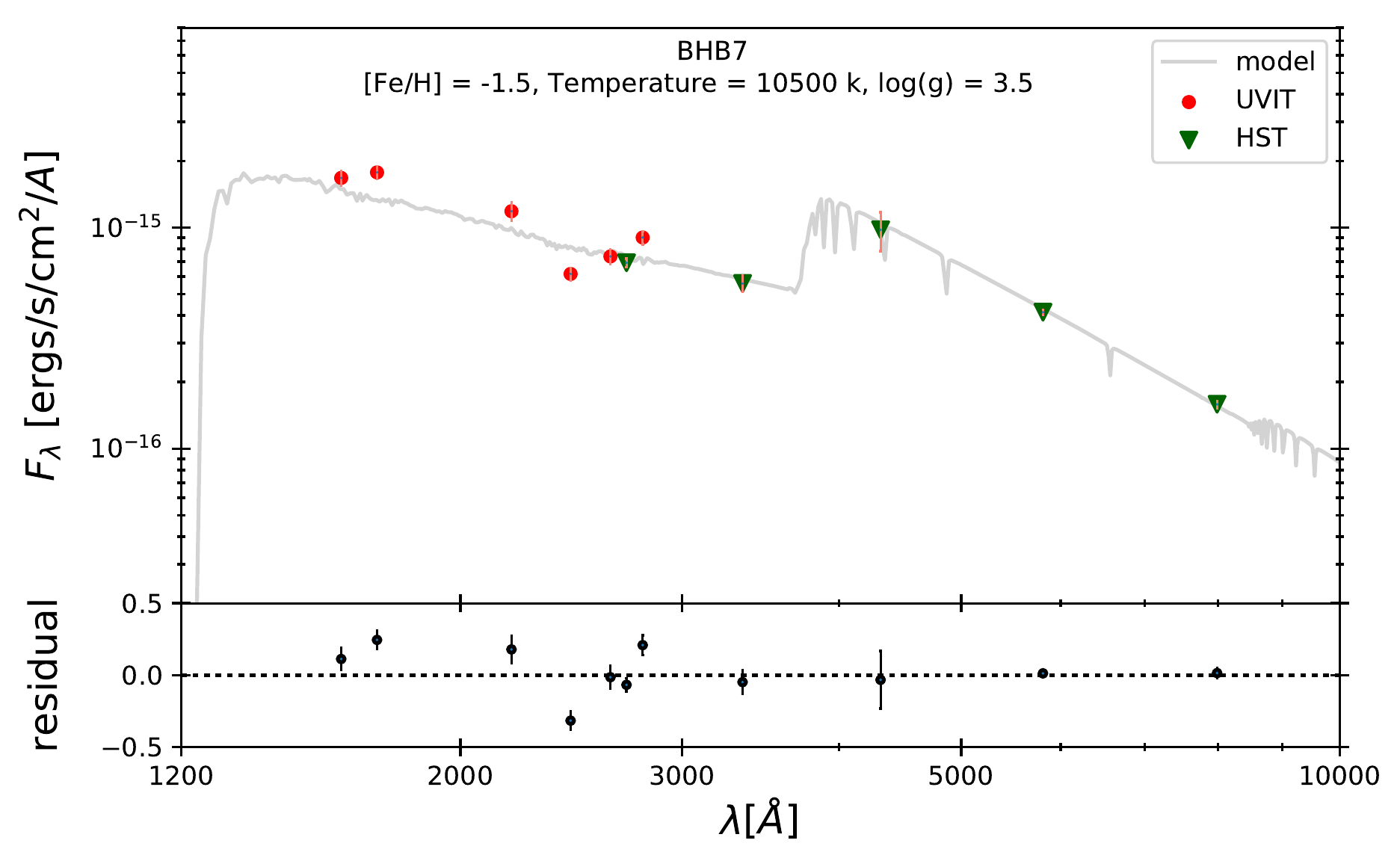}
  \end{subfigure}%
\begin{subfigure}{0.5\textwidth}
\centering
\includegraphics[height=8.0cm,width=9.0cm]{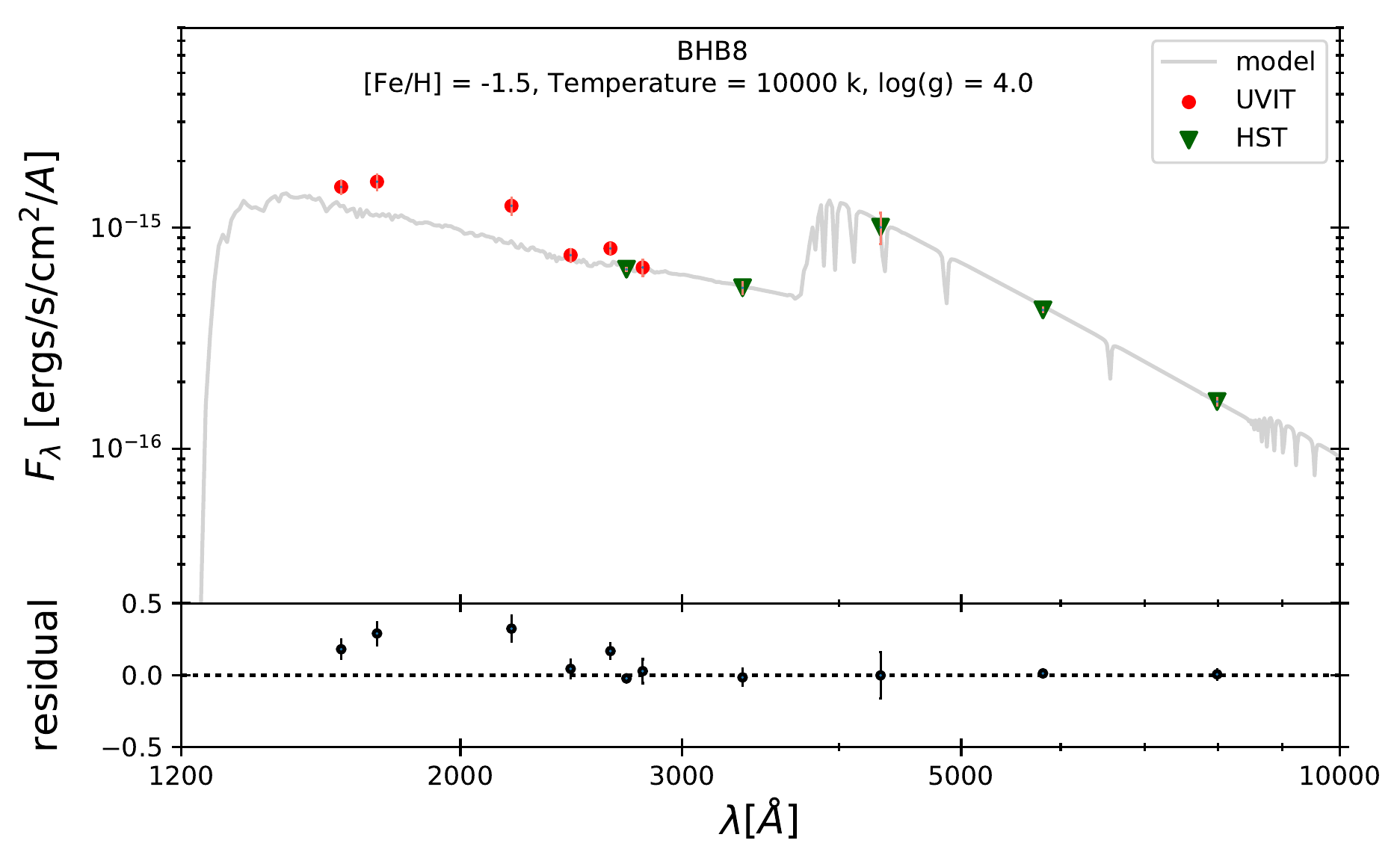}
\end{subfigure}
\begin{subfigure}{0.5\textwidth}
\centering
\includegraphics[height=8.0cm,width=9.0cm]{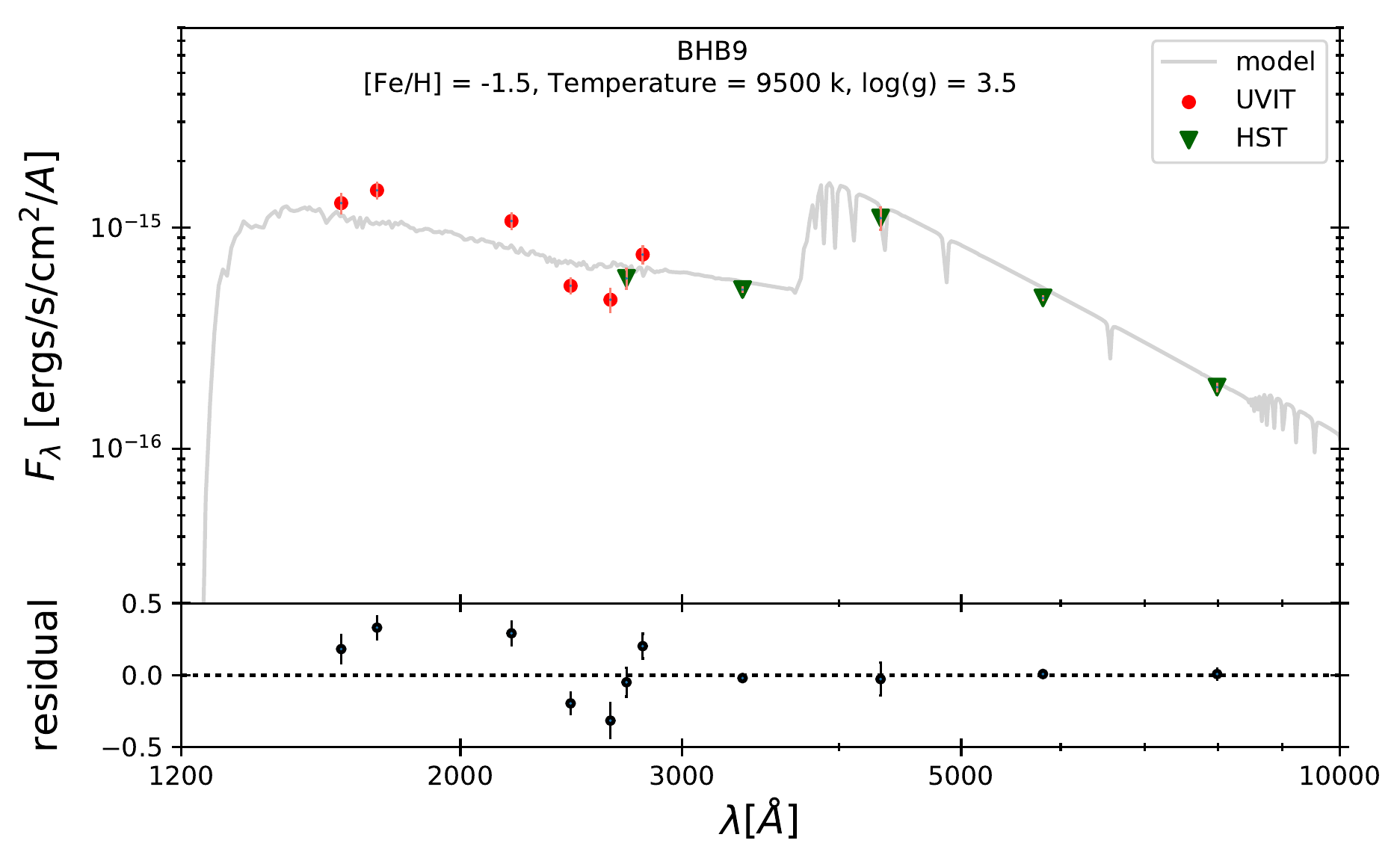}
\end{subfigure}%
\begin{subfigure}{0.5\textwidth}
\centering
\includegraphics[height=8.0cm,width=9.0cm]{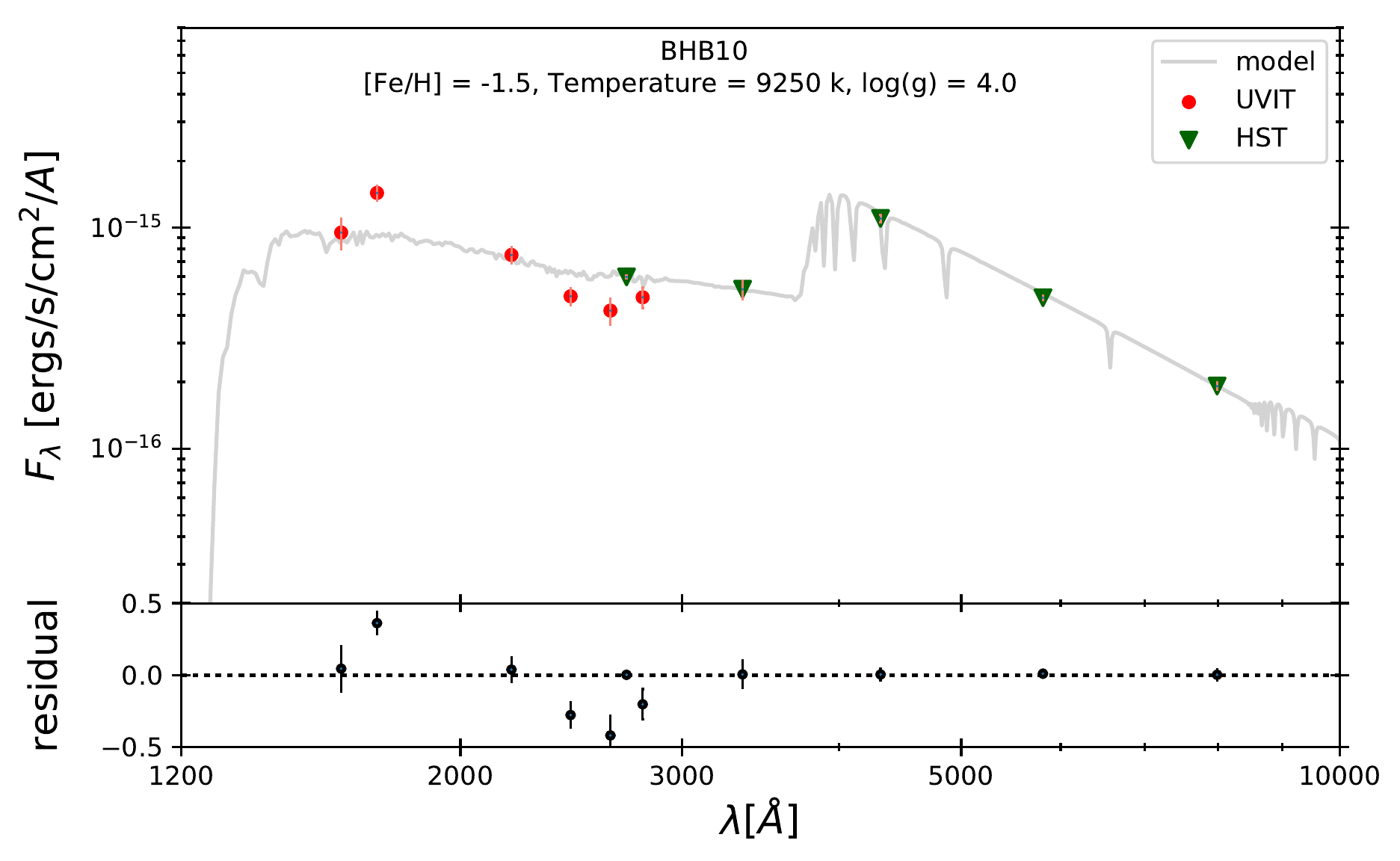}
\end{subfigure}
\begin{subfigure}{0.5\textwidth}
\centering
\includegraphics[height=8.0cm,width=9.0cm]{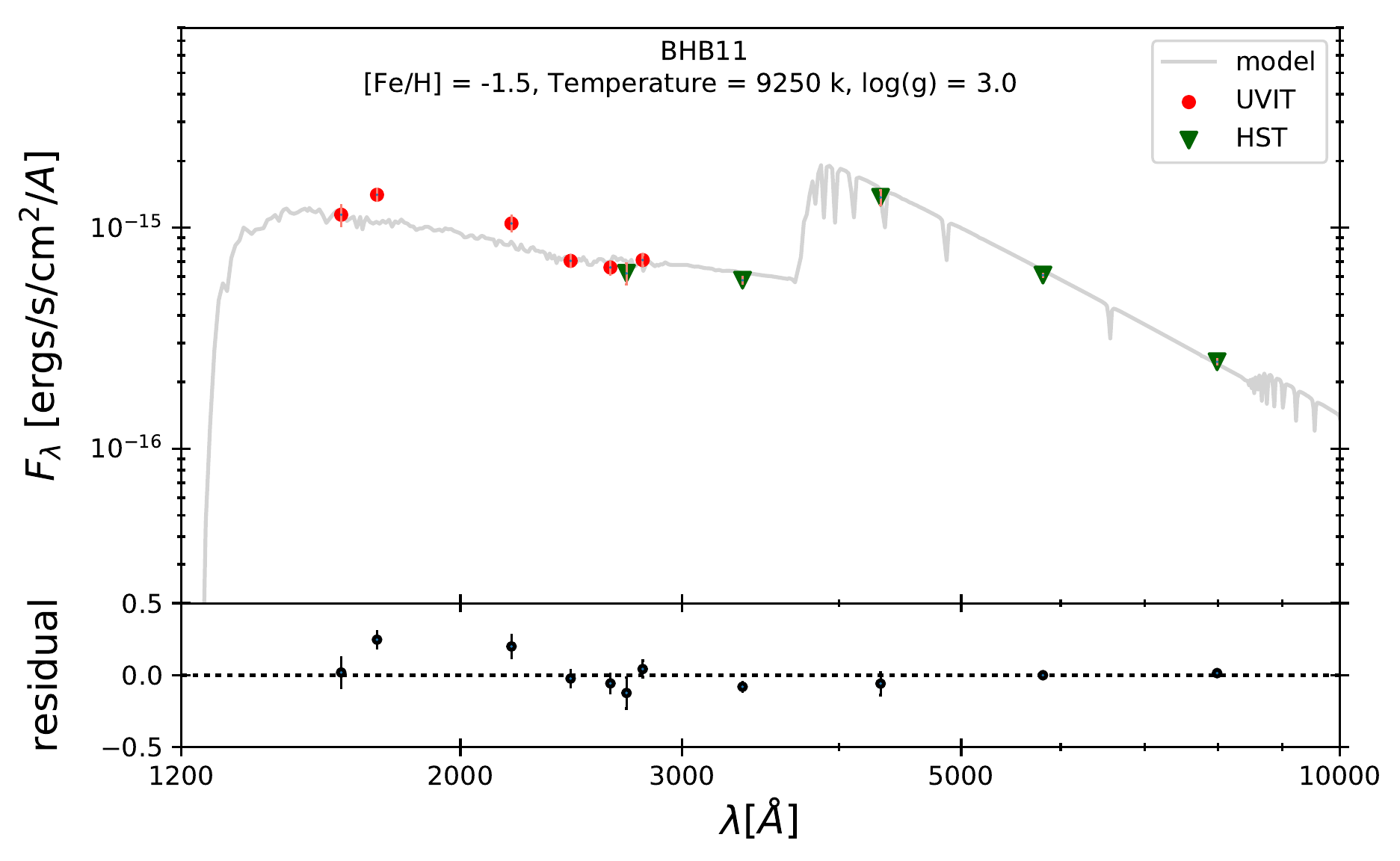}
\end{subfigure}%
\begin{subfigure}{0.5\textwidth}
\centering
\includegraphics[height=8.0cm,width=9.0cm]{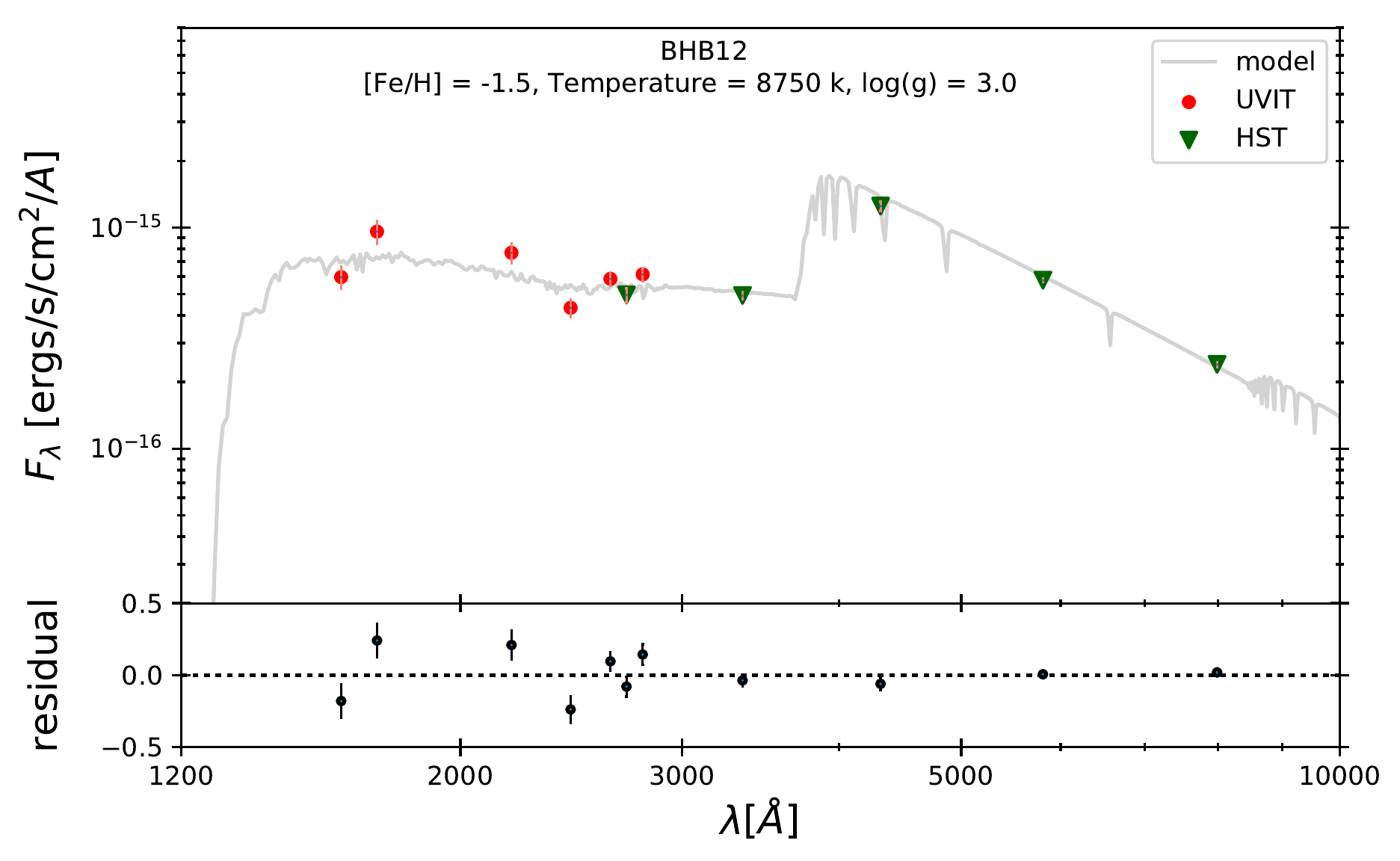}
\end{subfigure}
\caption{(Continued.)}
\label{sed2}
\end{figure*}
\renewcommand{\thefigure}{\arabic{figure}}

\renewcommand{\thefigure}{A\arabic{figure}}
\addtocounter{figure}{-1}
\begin{figure*}
\centering
\begin{subfigure}{0.5\textwidth}
\centering
\includegraphics[height=8.0cm,width=9.0cm]{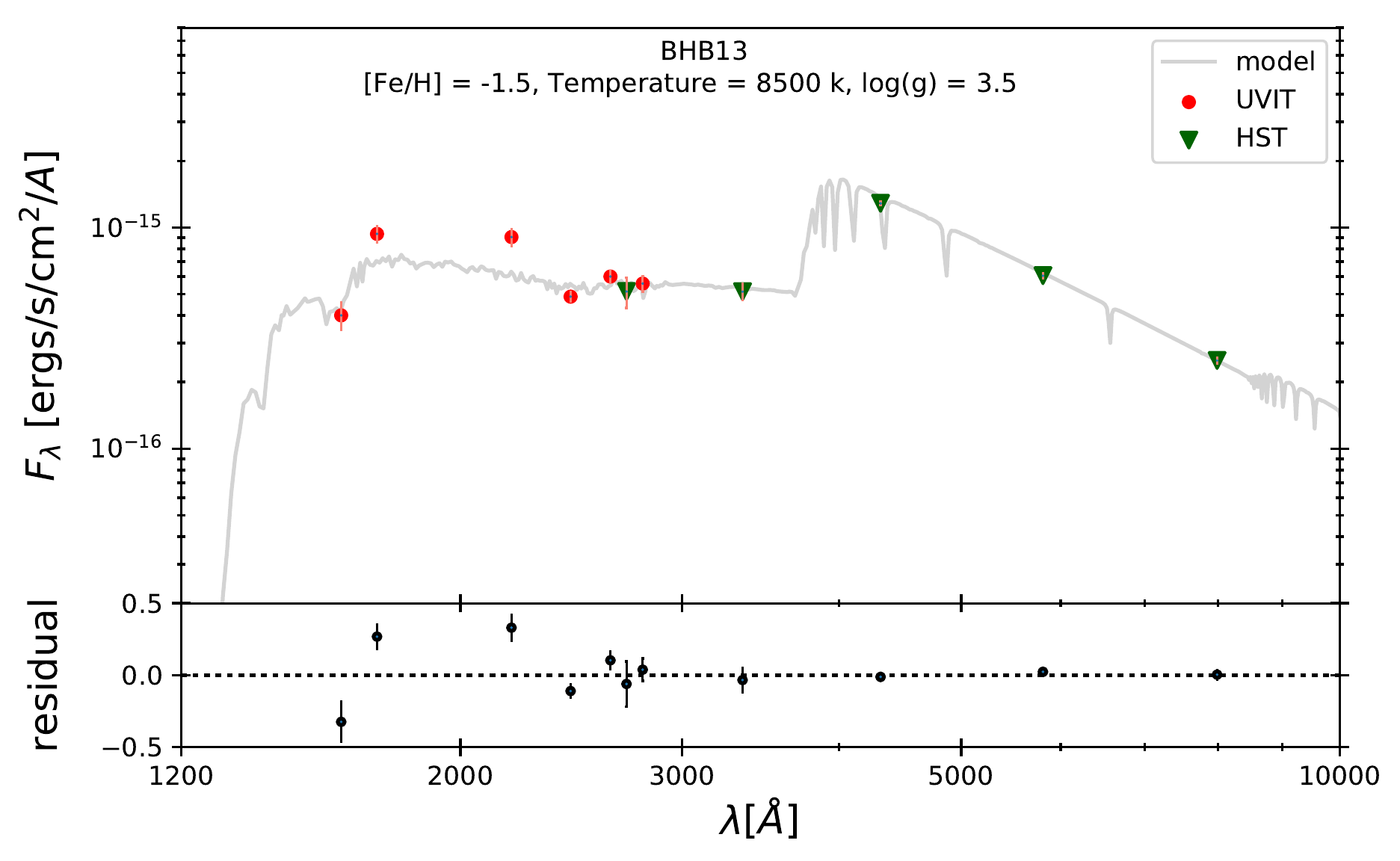}
\end{subfigure}%
\begin{subfigure}{0.5\textwidth}
\centering
\includegraphics[height=8.0cm,width=9.0cm]{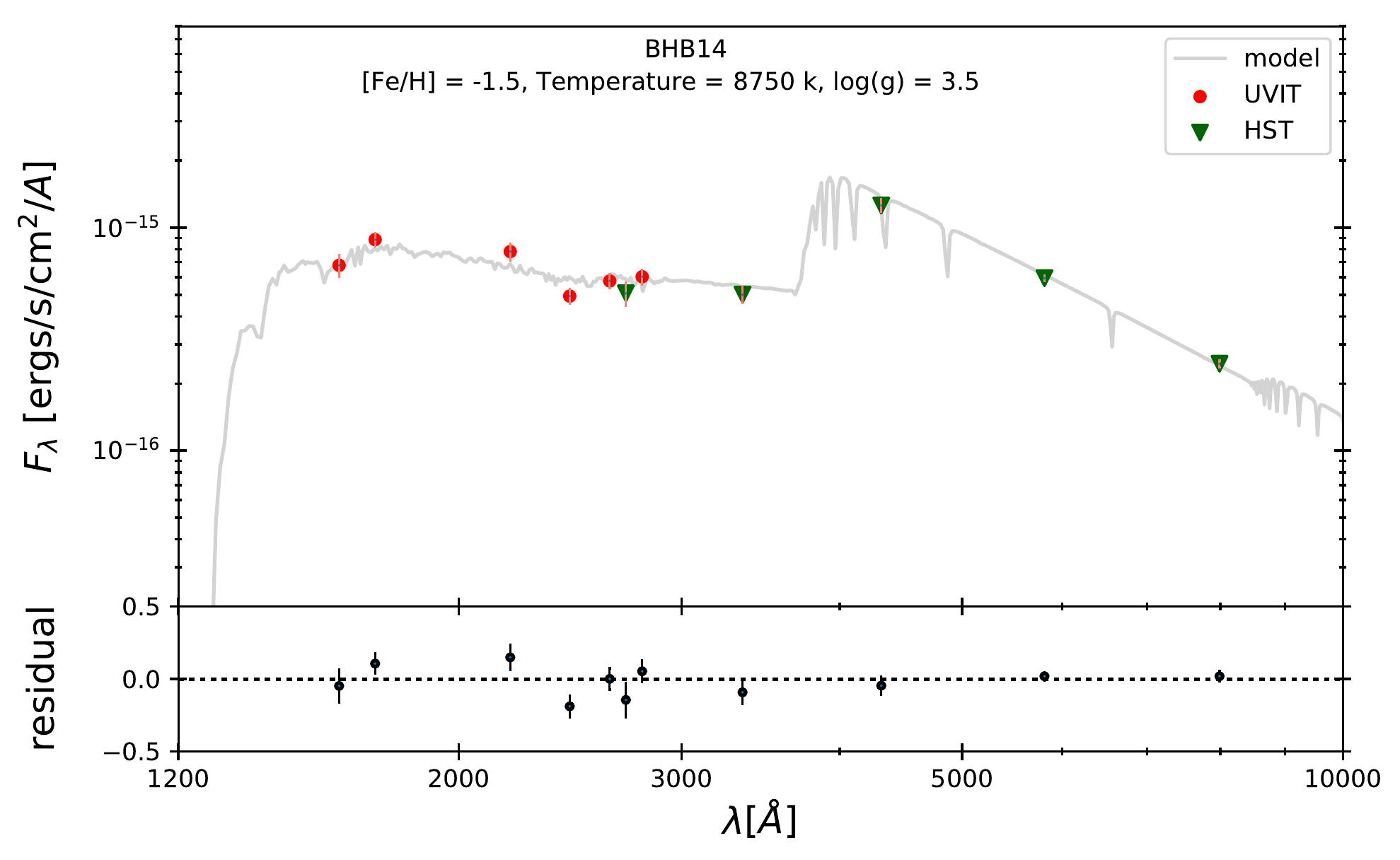}
\end{subfigure}
\begin{subfigure}{0.5\textwidth}
\centering
\includegraphics[height=8.0cm,width=9.0cm]{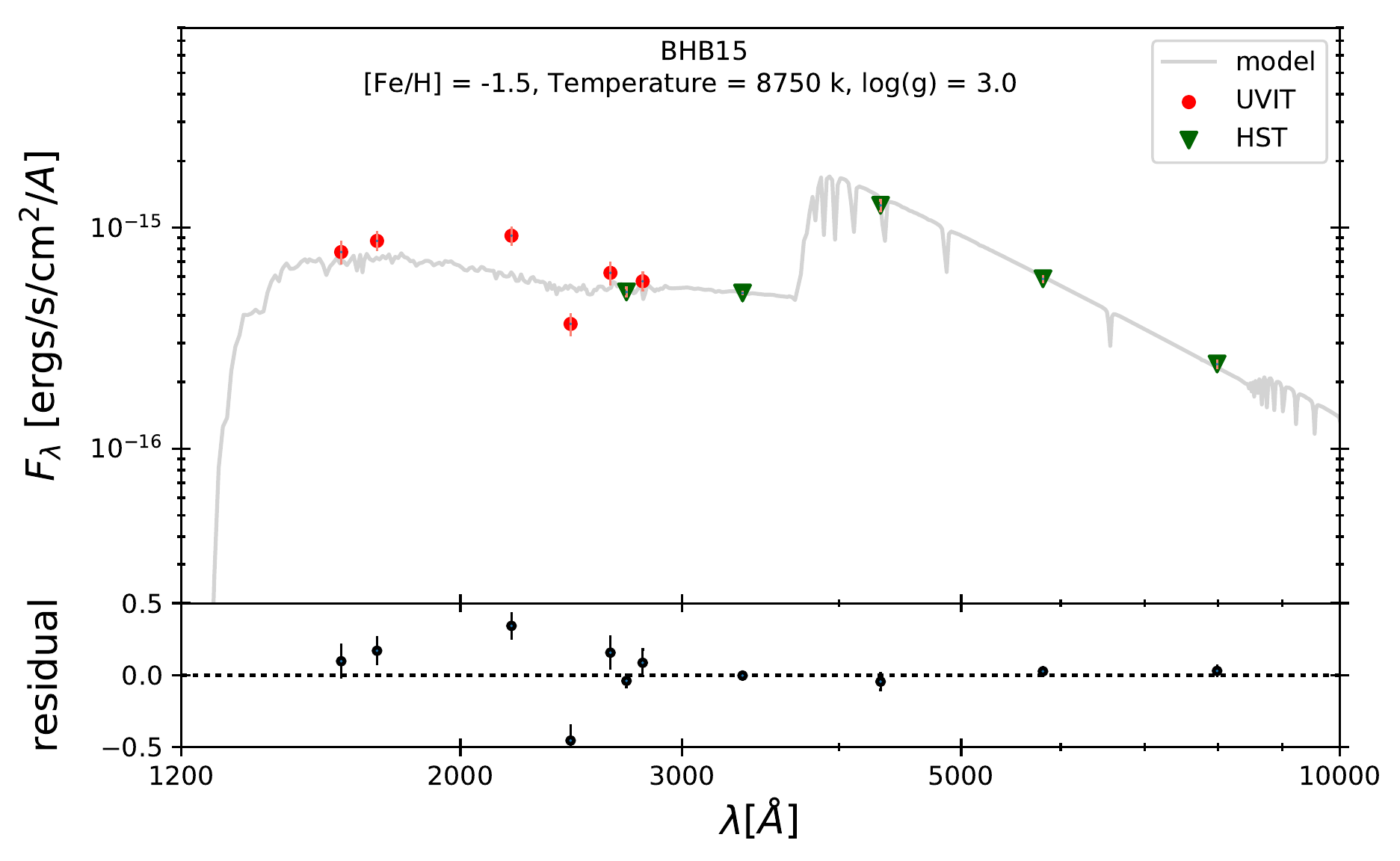}
\end{subfigure}%
\begin{subfigure}{0.5\textwidth}
\centering
\includegraphics[height=8.0cm,width=9.0cm]{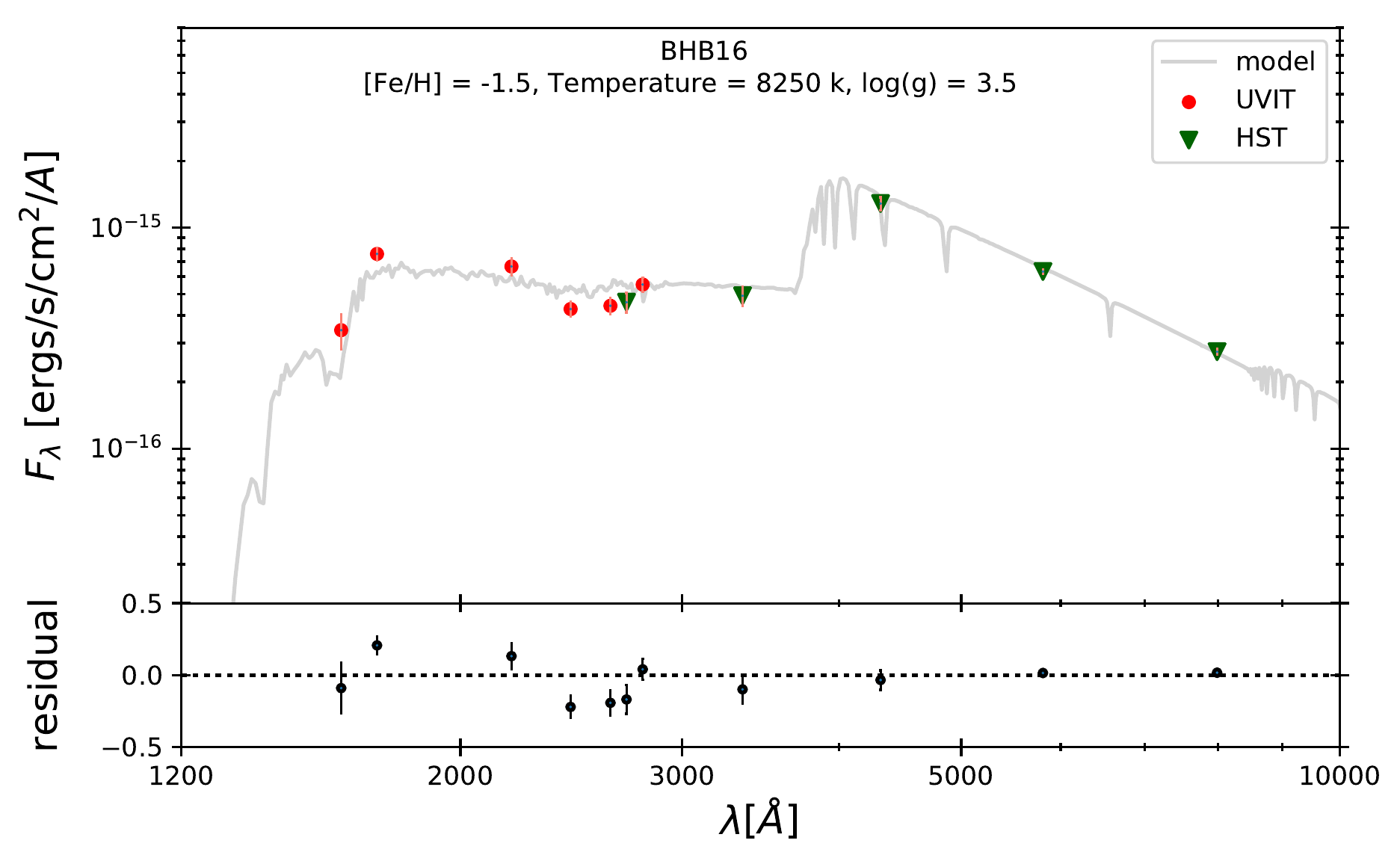}
\end{subfigure}
\begin{subfigure}{0.5\textwidth}
\centering
\includegraphics[height=8.0cm,width=9.0cm]{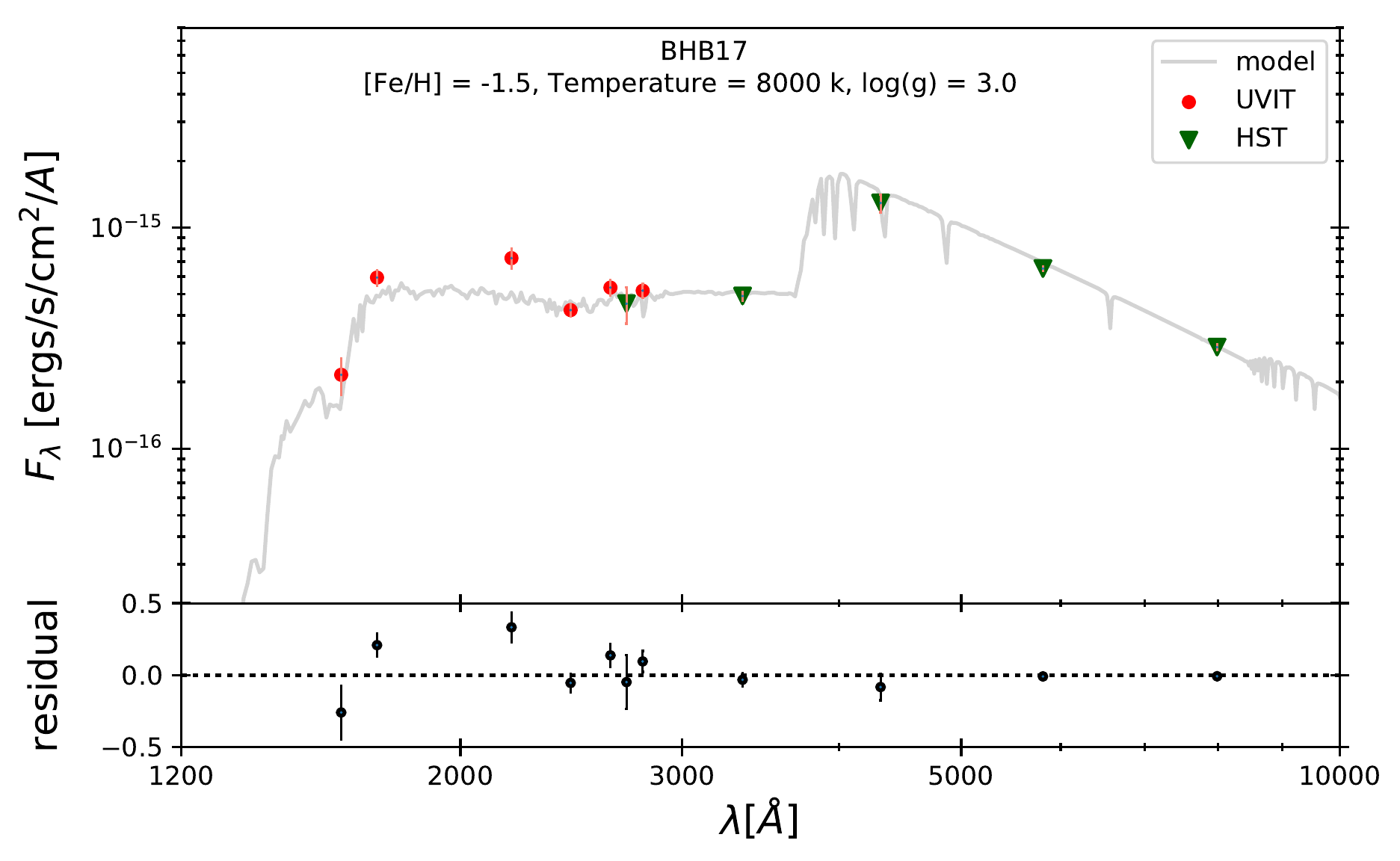}
\end{subfigure}%
\begin{subfigure}{0.5\textwidth}
\centering
\includegraphics[height=8.0cm,width=9.0cm]{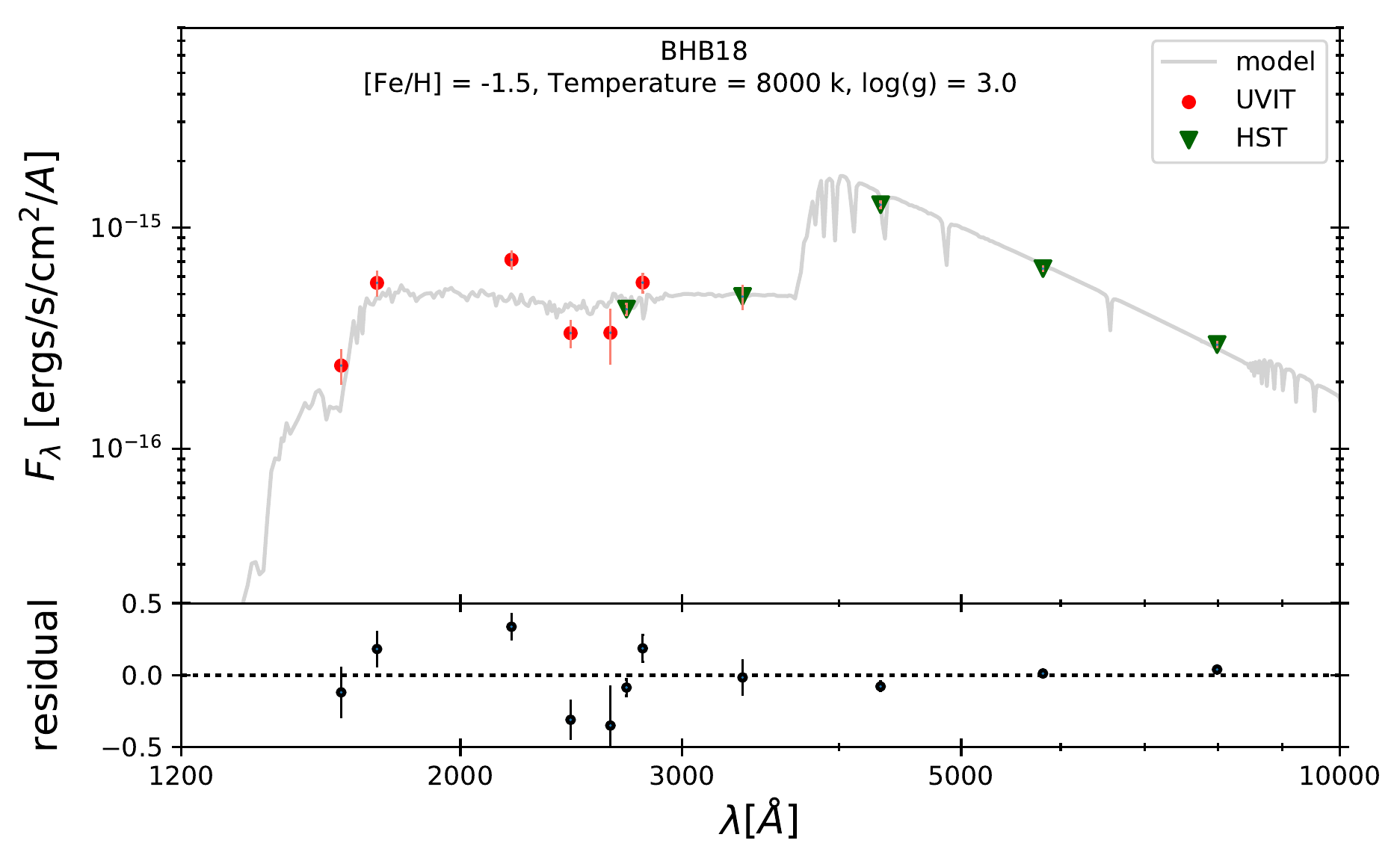}
\end{subfigure}
\caption{(Continued.)}
\label{sed3}
\end{figure*}
\renewcommand{\thefigure}{\arabic{figure}}

\renewcommand{\thefigure}{A\arabic{figure}}
\addtocounter{figure}{-1}
\begin{figure*}
\centering
\begin{subfigure}{0.5\textwidth}
\centering
\includegraphics[height=8.0cm,width=9.0cm]{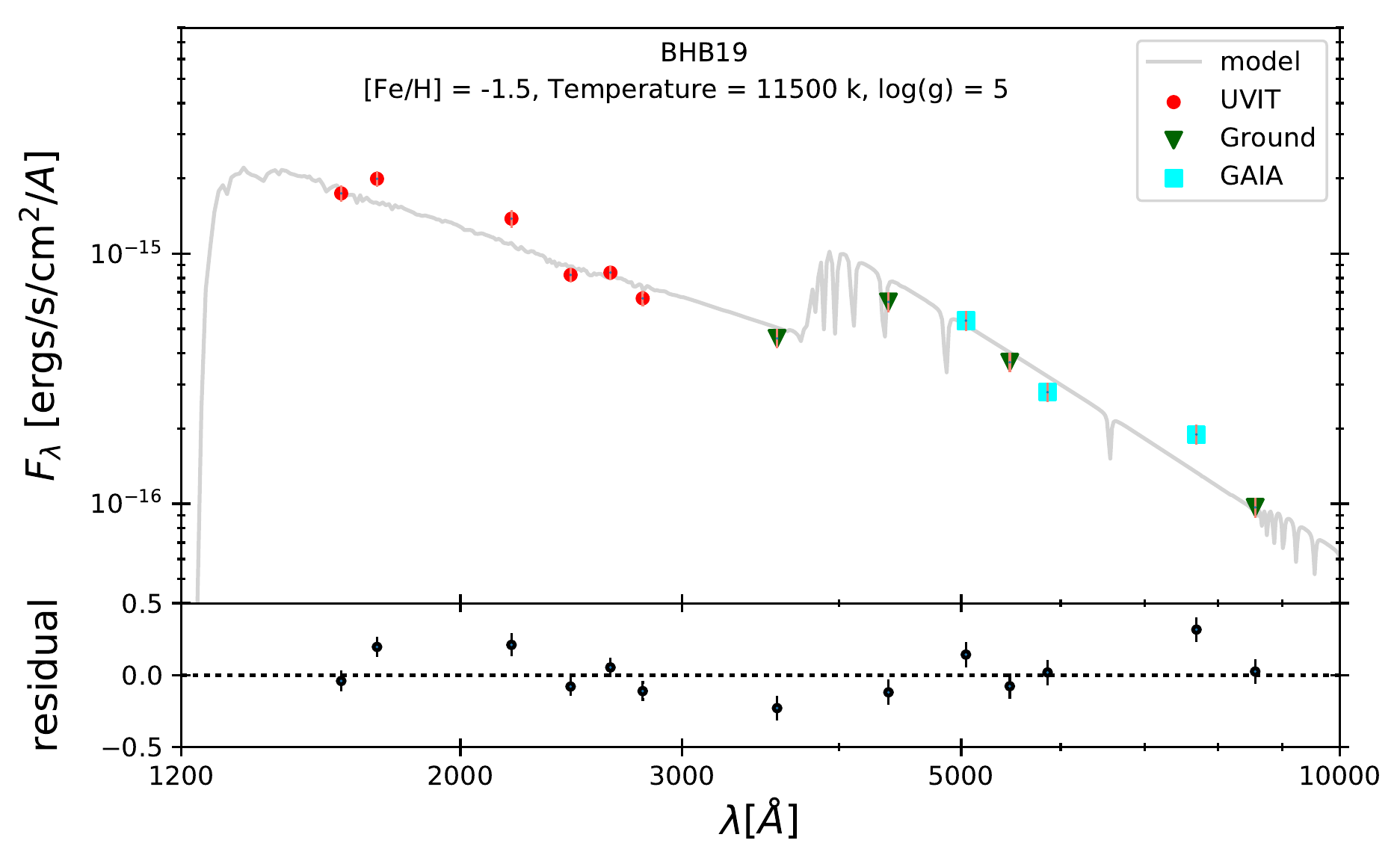}
\end{subfigure}%
\begin{subfigure}{0.5\textwidth}
\centering
\includegraphics[height=8.0cm,width=9.0cm]{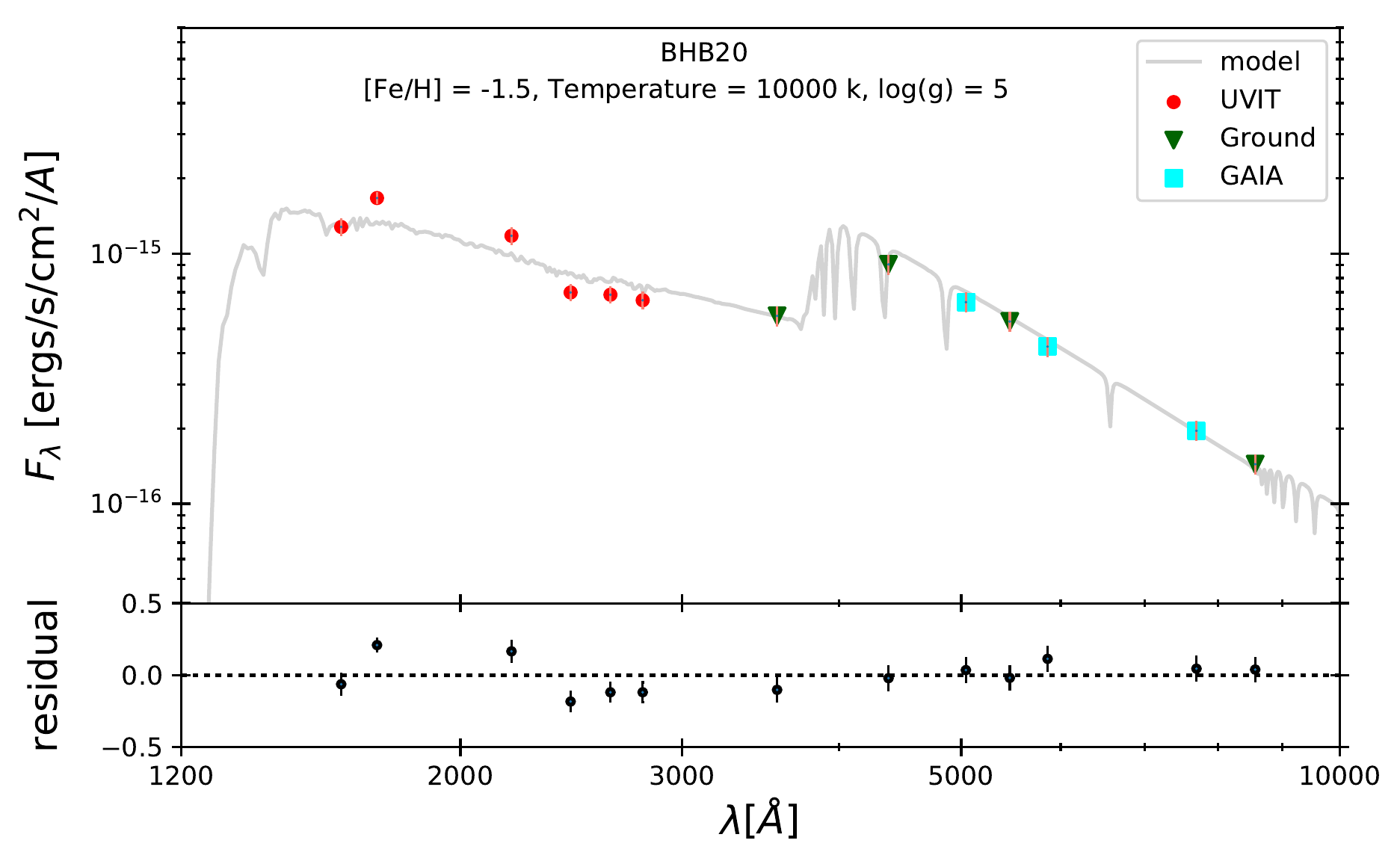}
\end{subfigure}
\begin{subfigure}{0.5\textwidth}
\centering
\includegraphics[height=8.0cm,width=9.0cm]{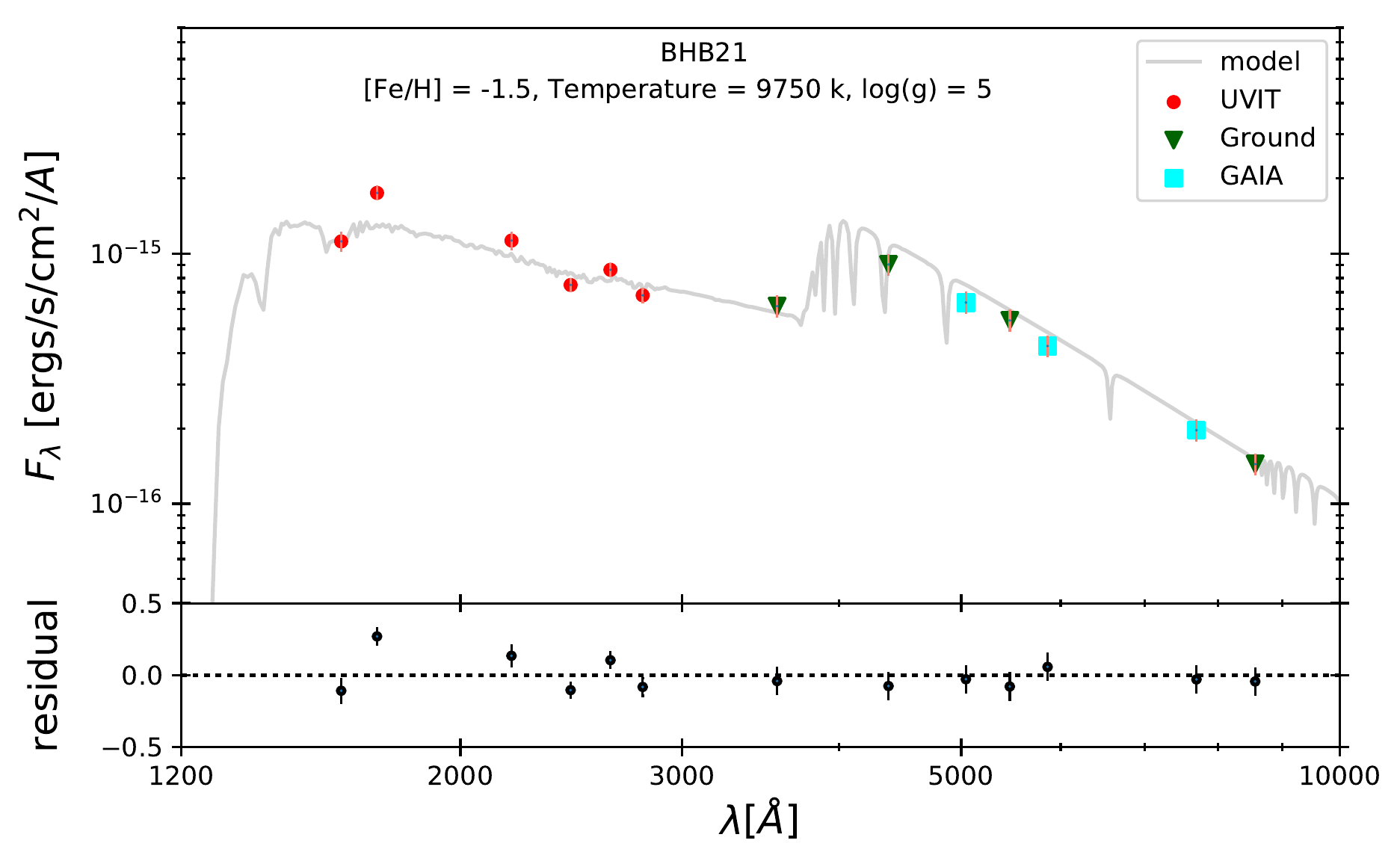}
\end{subfigure}%
\begin{subfigure}{0.5\textwidth}
\centering
\includegraphics[height=8.0cm,width=9.0cm]{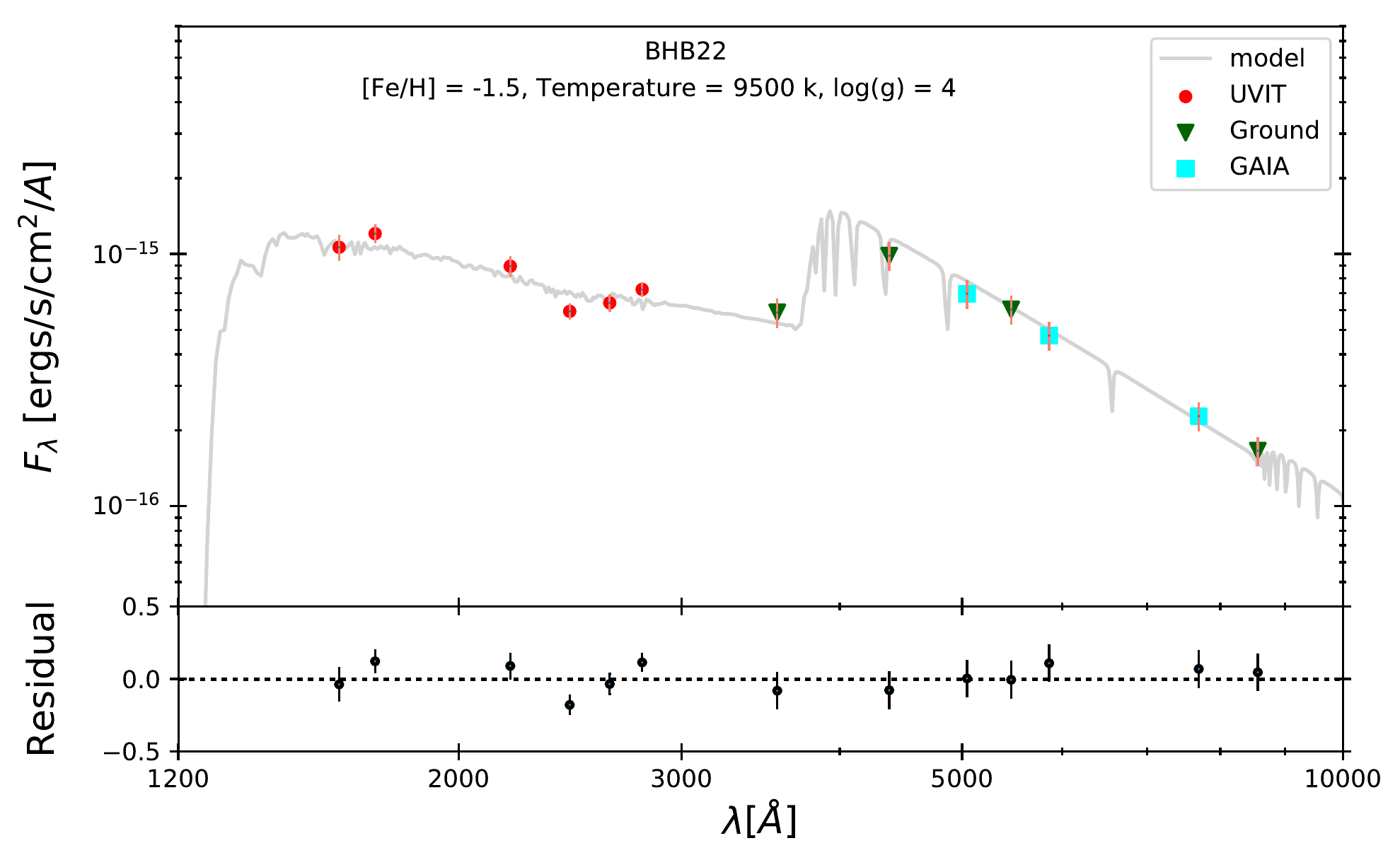}
\end{subfigure}
\begin{subfigure}{0.5\textwidth}
\centering
\includegraphics[height=8.0cm,width=9.0cm]{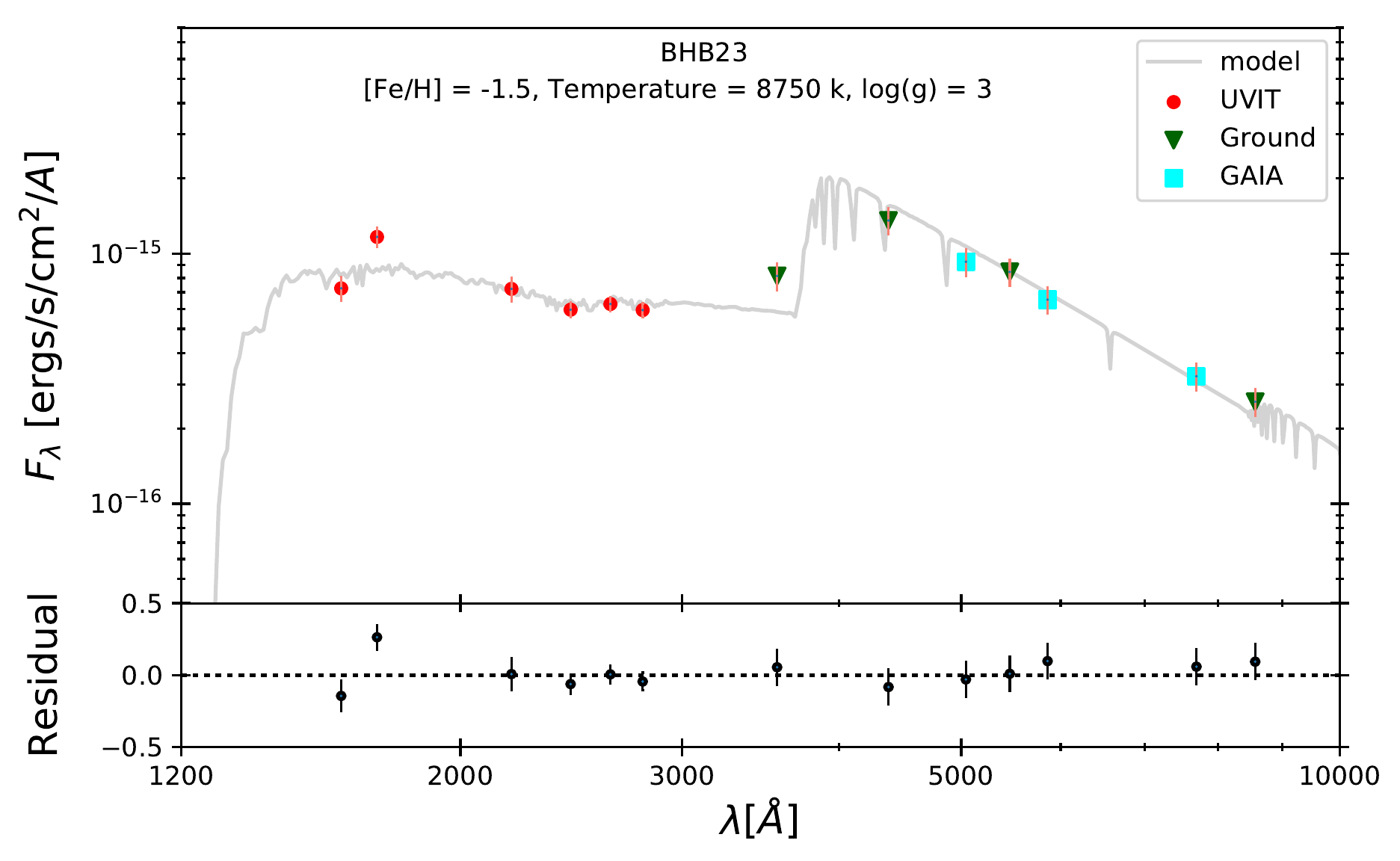}
\end{subfigure}%
\begin{subfigure}{0.5\textwidth}
\centering
\includegraphics[height=8.0cm,width=9.0cm]{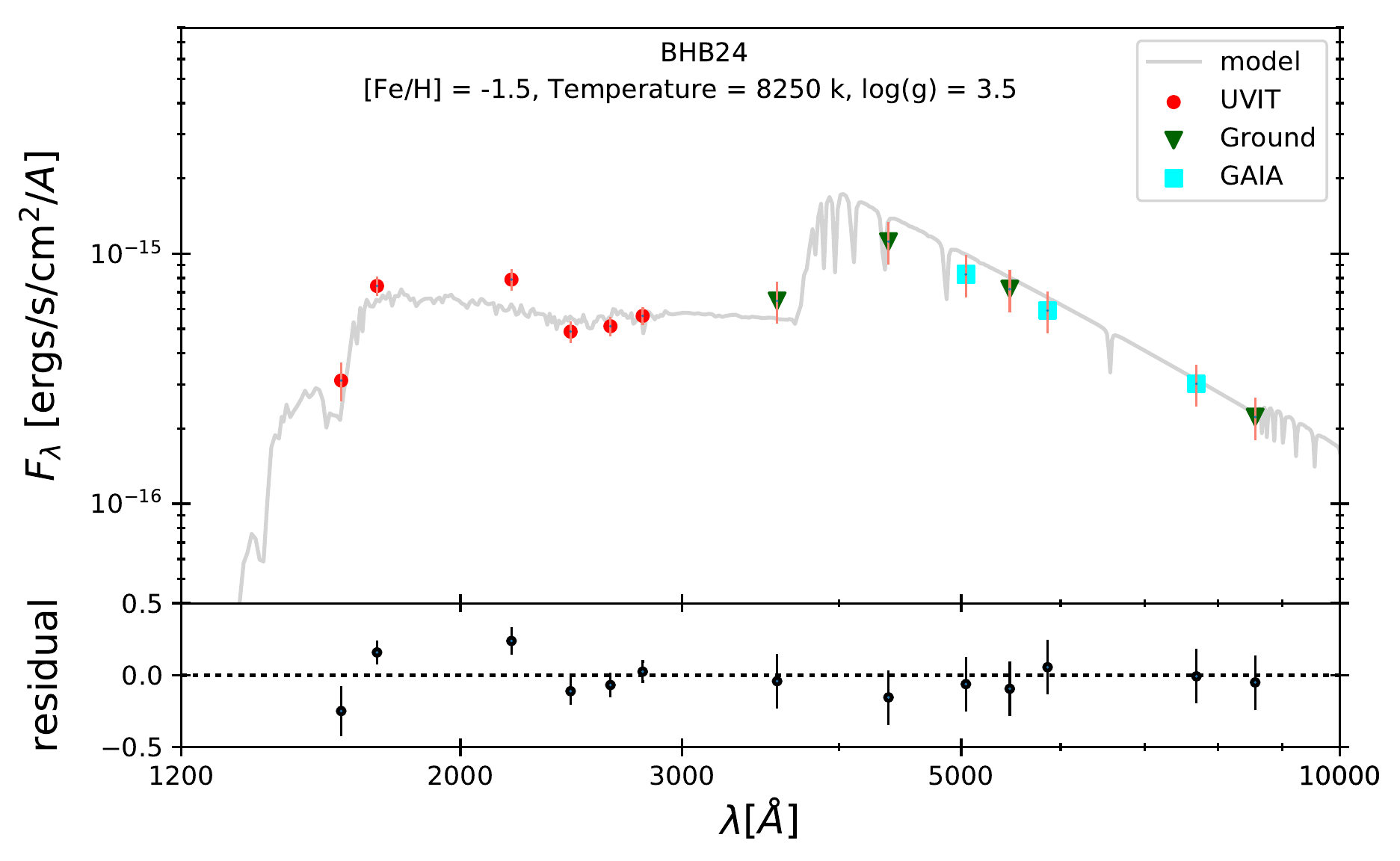}
\end{subfigure}
\caption{(Continued.)}
\label{sed4}
\end{figure*}
\renewcommand{\thefigure}{\arabic{figure}}


\bsp	
\label{lastpage}
\end{document}